\RequirePackage{fix-cm}
\documentclass[10pt,a4paper]{article}
\usepackage[margin=0.8in]{geometry}
\usepackage[T1]{fontenc}
\usepackage[utf8]{inputenc}
\usepackage{times,microtype}
\usepackage{amsmath,amssymb,bm,booktabs,tabularx,array,graphicx,float}
\usepackage{newtxmath}
\usepackage{tikz}
\usetikzlibrary{arrows.meta}
\usepackage{caption,titlesec,stfloats,multicol,ragged2e,needspace,etoolbox,placeins}
\usepackage[numbers,sort&compress]{natbib}
\usepackage[hidelinks,hyperfootnotes=false]{hyperref}
\hypersetup{pdftitle={Discrete Backbone Elasticity and Curvature-Dependent Torsion Statistics in Proteins},pdfauthor={Yiquan Wang, Mingyu Bian, Jiayuan Ma}}
\titleformat{\section}{\large\bfseries\justifying\hyphenpenalty=10000\exhyphenpenalty=10000\setlength{\parindent}{0pt}}{}{0pt}{}
\titleformat{\subsection}{\normalsize\bfseries\justifying\hyphenpenalty=10000\exhyphenpenalty=10000\setlength{\parindent}{0pt}}{}{0pt}{}
\titlespacing*{\section}{0pt}{2.2ex plus .4ex minus .2ex}{.9ex}
\titlespacing*{\subsection}{0pt}{1.6ex plus .3ex minus .2ex}{.6ex}
\pretocmd{\section}{\Needspace{7\baselineskip}}{}{}
\pretocmd{\subsection}{\Needspace{6\baselineskip}}{}{}
\renewenvironment{abstract}{\par\noindent\textbf{Abstract}\par\smallskip\noindent\ignorespaces}{\par}
\graphicspath{{figures/}}
\title{\bfseries Discrete Backbone Elasticity and Curvature-Dependent Torsion Statistics in Proteins}
\author{Yiquan Wang\textsuperscript{1,2,3,*}, Mingyu Bian\textsuperscript{2,4}, Jiayuan Ma\textsuperscript{5}}
\date{}
\makeatletter
\renewcommand{\maketitle}{%
\begin{center}
\vspace*{4pt}
{\LARGE\@title\par}
\vspace{8pt}
{\normalsize\@author\par}
\vspace{3pt}
{\fontsize{9}{10}\selectfont
\textsuperscript{1}College of Mathematics and System Sciences, Xinjiang University, Urumqi 830046, Xinjiang, China.\\[1pt]
\textsuperscript{2}Xinjiang Key Laboratory of Biological Resources and Genetic Engineering, College of Life Science and Technology, Xinjiang University, Urumqi 830049, Xinjiang, China.\\[1pt]
\textsuperscript{3}Institute of Systems and Physical Biology, Shenzhen Bay Laboratory, Shenzhen 518132, China.\\[1pt]
\textsuperscript{4}School of Intelligence Science and Technology, Xinjiang University, Urumqi 830017, Xinjiang, China.\\[1pt]
\textsuperscript{5}School of Life Sciences, Inner Mongolia University, Hohhot 010070, China.\\[3pt]
\textsuperscript{*}Corresponding author: Yiquan Wang (\href{mailto:ethan@stu.xju.edu.cn}{ethan@stu.xju.edu.cn})\par}
\end{center}
}
\makeatother

\begin{document}
\twocolumn[{
\begin{@twocolumnfalse}
\maketitle
\vspace{3pt}
\begin{abstract}
Coarse-grained models of protein conformational dynamics use $\mathrm{C}_\alpha$ geometry to describe the probabilities of regular helices, ordered turns and disordered conformations. To test how much of this order arises from local chain elasticity, we connect discrete Hasimoto frame transport to neighboring bending mismatch, deriving a curvature-product torsion interaction and a six-parameter generalized von Mises density. Fitted on 194 chains from the ATLAS molecular dynamics database and tested on 1744 sequence-nonredundant proteins, this baseline reduces negative log-density loss from 1.490 to 1.160 nats, torsion-sign error from 12.26 to 2.15 percentage points, and helix-fraction error from 21.84 to 10.41 points. Ablations and circular-moment projections against reference generative models motivate extensions to anisotropic stiffness and two partition-function-weighted elastic states, further lowering test loss to 0.790 nats and helix error to 1.65 points. The fixed hierarchy also improves predictions across 72 trajectories of peptides $\mathrm{Fs}$ and $(\mathrm{AAQAA})_3$ and miniprotein $\mathrm{HP35}$ spanning varied starting conformations and force fields. Remaining angular errors in these trajectories are associated with atomic context at shared bends, as bend-matched peptide frames adopt opposite torsion signs with distinct hydrogen-bond patterns and a proline-containing $\mathrm{HP35}$ turn remains at negative torsion. Introducing a bilinearly coupled $\mathrm{C}_\alpha$--$\mathrm{N}$ orientation angle largely reparameterizes the existing bend-dependent harmonics, with an empirical $L^2$ Gram projection placing 96.7\% to 99.6\% of its added coefficient-function squared norm within the original six-function span. These results link discrete Hasimoto elasticity to statistical backbone potentials and clarify the scope and limits of local $\mathrm{C}_\alpha$ geometry.
\end{abstract}
\vspace{6pt}
\end{@twocolumnfalse}
}]

\section{Introduction}
\label{sec:introduction}

Protein conformational dynamics contributes to molecular recognition, catalysis and allosteric regulation through changes in the populations and accessibility of functional states~\cite{henzler2007dynamic,wankowicz2025advances}. Characterizing these processes benefits from a statistical description of the conformations available to proteins. Coarse-grained models reduce atomic detail to explore these conformational ensembles at a lower computational cost~\cite{kmiecik2016coarse}, and recent models learned from atomistic data extend this approach to conformational landscapes across protein sequences~\cite{charron2025navigating}. In representations based on $\mathrm{C}_\alpha$ atoms, effective interactions aim to capture the conformational preferences of protein backbones across regular helices, ordered turns and disordered regions.

Four consecutive $\mathrm{C}_\alpha$ atoms form a local unit with two adjacent bending angles and one signed torsion angle. Both helical and nonhelical units can have concentrated torsion distributions, so a departure from helical geometry does not by itself quantify local disorder. Existing models retain different sources of conformational information. The five-parameter Fisher--Bingham hidden Markov model, FB5-HMM, combines continuous $\mathrm{C}_\alpha$ angular distributions with correlated hidden conformational states~\cite{hamelryck2006sampling}. The toroidal dynamic Bayesian network, TorusDBN, models atomic backbone dihedral angles and peptide-bond states conditional on sequence~\cite{boomsma2008generative}. Lifson--Roig theory represents helix--coil cooperativity through nucleation and propagation weights for residue-level helical states~\cite{lifson1961theory}.

Local angular preferences also affect the organization of longer chain segments. Multitorsional potentials in the united-residue (UNRES) force field couple successive virtual torsions and improve the modeled geometry of helices and their connecting regions~\cite{petrusevich2025recapturing}. In static protein structures, changes in torsion sign between adjacent backbone positions are associated with spatial variations in curvature and with nonlocal contacts~\cite{wang2026instability}. These findings motivate testing a direct elastic interaction between neighboring bends as a source of local torsional order.

The Hasimoto representation combines bending amplitudes with accumulated torsional phases~\cite{hasimoto1972soliton,molkenthin2011discrete}, while discrete frame transport relates these variables to backbone rotations~\cite{hu2011discrete,danielsson2010gauge}. Existing Hasimoto formulations describe static backbone geometry through discrete solitons and spectral boundaries~\cite{wang2026descriptive,wang2026spectral}, but they provide deterministic curves rather than statistical distributions for thermal conformational ensembles. We bridge this geometric representation to molecular dynamics by treating the rotational mismatch between neighboring bending generators in $\mathfrak{so}(3)$ as an elastic energy. In the shared normal plane, this mismatch yields an analytical Hamiltonian where the curvature product modulates torsional stiffness and the Lie commutator determines chirality. This transforms a deterministic geometric identity into a rigorous statistical mechanical framework, generating a continuous six-parameter conditional probability density for fluctuating protein trajectories.

We evaluate this local elastic interaction on 1744 ATLAS protein chains after fitting on a separate set of 194 chains~\cite{vander2024atlas}. Published FB5-HMM and TorusDBN parameters provide reference predictions, and standard Lifson--Roig theory is evaluated using its residue-level helical states. The six-parameter model improves direction probabilities over a curvature-independent density while leaving helical peaks too broad. To address this limitation, we examine two elastic states with different bending preferences and anisotropic responses whose relative weights combine bend energies with the angular partition function integrating the torsional Boltzmann weight of each state. Fixed ATLAS parameters are then tested on 72 controlled trajectories of the 21-residue alanine-rich Fs peptide, the 15-residue peptide $(\mathrm{AAQAA})_3$ and the 35-residue villin headpiece HP35. Four starting conformations, three random repeats and two force-field protocols test transfer across sampled backgrounds. Geometry-matched atomic comparisons and elimination of an atomic orientation around the central virtual bond relate the remaining angular errors to structural context.

\section{Theory}
\label{sec:theory}
\label{sec:geometry}

For consecutive C$_\alpha$ positions $\mathbf r_i$, define bond lengths $\ell_i=|\mathbf r_{i+1}-\mathbf r_i|$, tangents $\mathbf t_i=(\mathbf r_{i+1}-\mathbf r_i)/\ell_i$ and turns $\kappa_i=\arccos(\mathbf t_i\cdot\mathbf t_{i+1})$. At regular bends, $0<\kappa_i<\pi$, the unit binormal is $\mathbf b_i=\widehat{\mathbf t_i\times\mathbf t_{i+1}}$ and the signed torsion is
\begin{equation}
\label{eq:torsion}
\tau_i=\operatorname{atan2}\!\left[\mathbf t_{i+1}\cdot(\mathbf b_i\times\mathbf b_{i+1}),\mathbf b_i\cdot\mathbf b_{i+1}\right].
\end{equation}
A four-C$_\alpha$ unit has normalized bends $x_L=\kappa_L/\pi$, $x_R=\kappa_R/\pi$ and product $r=x_Lx_R$. Its two binormals lie in the plane normal to the shared tangent $\mathbf t$, with dot and directed cross products $\cos\tau$ and $\sin\tau$.

The normalized generators of the minimal bending rotations are $\mathbf a_L=x_L\mathbf b_L$ and $\mathbf a_R=x_R\mathbf b_R$. We assume a quadratic mismatch and a directed pairing,
\begin{equation}
\label{eq:elastic-metrics}
\begin{aligned}
 E_{\mathrm{pair}}&=\frac K2|\mathbf a_R-\mathbf a_L|^2-J_s\mathbf t\cdot(\mathbf a_L\times\mathbf a_R)\\
 &=\frac K2(x_L^2+x_R^2)-r(K\cos\tau+J_s\sin\tau).
\end{aligned}
\end{equation}
Viewed as elements of the rotation Lie algebra $\mathfrak{so}(3)$ in the plane normal to $\mathbf t$, $\mathbf a_L\cdot\mathbf a_R=r\cos\tau$ is an invariant inner product proportional to the negative Killing form and $\mathbf t\cdot(\mathbf a_L\times\mathbf a_R)=r\sin\tau$ is the projection of the Lie commutator along $\mathbf t$. The commutator supplies the directed geometric pairing; a nonzero chiral coefficient $J_s$ introduces the parity-breaking energetic preference. Writing $K\cos\tau+J_s\sin\tau=\tilde K\cos(\tau-\tau_0)$ with $\tilde K=\sqrt{K^2+J_s^2}$ defines the preferred phase $\tau_0=\operatorname{atan2}(J_s,K)$ between neighboring bending planes ($\tau_0=49.8^\circ$ in the ATLAS fit, close to the canonical right-handed $\alpha$-helical torsion). At fixed bends, the diagonal term cancels from the normalized torsion density.

\begin{figure*}[tp]
\centering
\includegraphics[width=\textwidth]{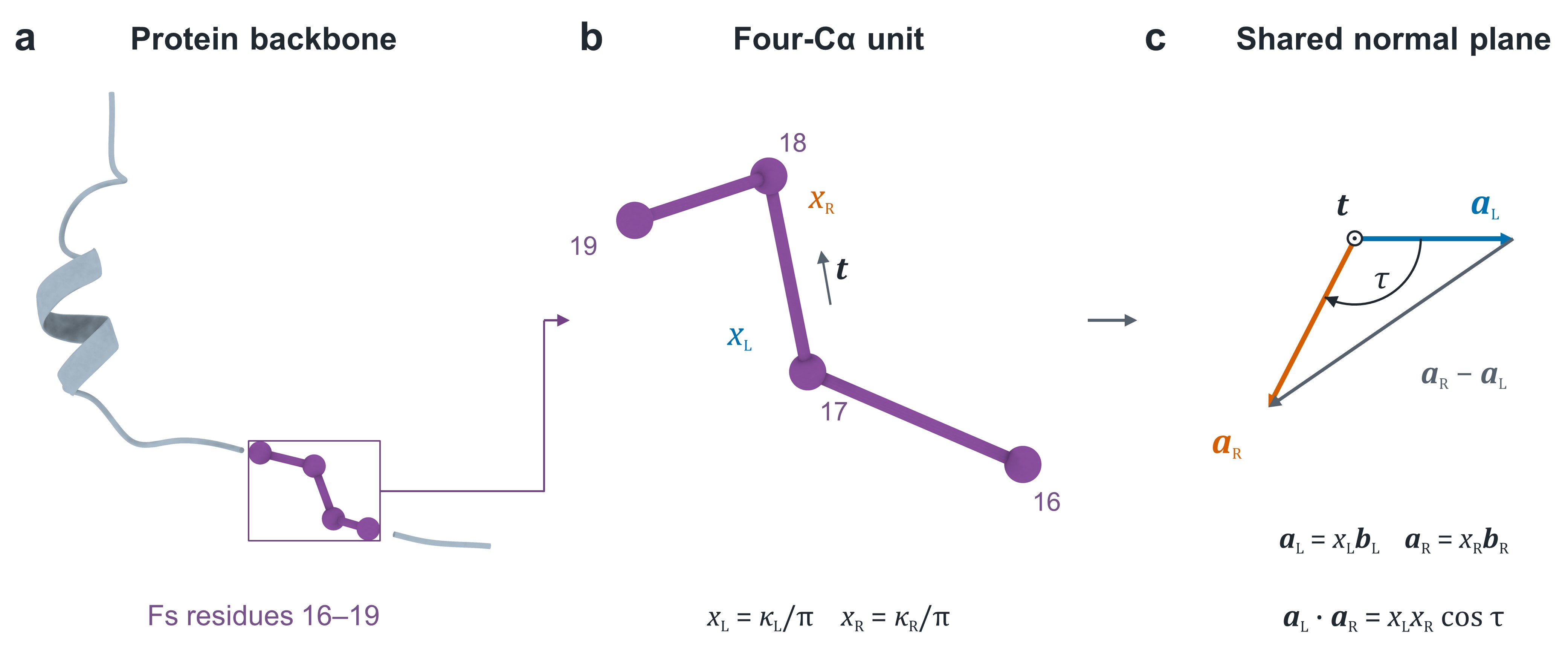}
\caption{Geometry of neighboring backbone bends. Panel a shows an Fs peptide conformation with residues 16--19 highlighted in purple. Panel b shows the corresponding four-C$_\alpha$ unit with normalized bends $x_L$ and $x_R$ and central tangent $\mathbf t$. Panel c expresses the bending generators in their shared normal plane, relating the curvature product and signed torsion through $\mathbf a_L\cdot\mathbf a_R=x_Lx_R\cos\tau$.}
\label{fig:structures}
\end{figure*}

Discrete frame transport gives the corresponding Hasimoto construction. For the column frame $F_i=(\mathbf n_i,\mathbf b_{i-1},\mathbf t_i)$ with $\mathbf n_i=\mathbf b_{i-1}\times\mathbf t_i$ and right-handed coordinate rotations $R_a$,
\begin{equation}
\label{eq:frame-transport}
 F_{i+1}=F_iR_3(\tau_{i-1})R_2(\kappa_i).
\end{equation}
With Pauli matrices $\sigma_a$, the corresponding $\mathrm{SU}(2)$ lift is $U_i=e^{-i\tau_{i-1}\sigma_3/2}e^{-i\kappa_i\sigma_2/2}$. Defining the discrete Hasimoto complex field $\zeta_i=x_i e^{i\varphi_i}$ with $x_i=\kappa_i/\pi$ and accumulated phase difference $\varphi_{i+1}-\varphi_i=\tau_i$ yields the adjacent-difference identity
\begin{equation}
\label{eq:hasimoto-mismatch}
 |\zeta_{i+1}-\zeta_i|^2=x_i^2+x_{i+1}^2-2x_ix_{i+1}\cos\tau_i.
\end{equation}
Frame transport uses the torsion about $\mathbf t_i$, whereas the Hasimoto phase difference pairs bends about $\mathbf t_{i+1}$. Equation~\eqref{eq:hasimoto-mismatch} expresses the isotropic bend mismatch as a Hasimoto phase-difference interaction, while the constant $\mathrm{U}(1)$ transport link $W_i=e^{i\tau_0}$ combines both angular terms into $\tfrac12\tilde K|\zeta_{i+1}-e^{i\tau_0}\zeta_i|^2$ up to the bend-only term $\tfrac12(K-\tilde K)(x_i^2+x_{i+1}^2)$, which cancels from the normalized torsion density at fixed bends.

We represent the remaining angular background by its first two circular harmonics. For torsion angle $t\in[-\pi,\pi)$ conditional on a scalar bending variable $s$, the maximum-entropy generalized von Mises density~\cite{gatto2007generalized} takes the form $p(t\mid s)=Z(s)^{-1}e^{g(t,s)}$, with normalization $Z(s)=\int_{-\pi}^{\pi}e^{g(t,s)}dt$ and angular potential
\begin{equation}
\label{eq:angular-density}
\begin{aligned}
 g(t,s)&=(a_0+a_1s)\cos t+(b_0+b_1s)\sin t\\
 &\quad+c_0\cos2t+d_0\sin2t.
\end{aligned}
\end{equation}
The single-state elastic model sets $s=r$, with $a_1=\beta K$ and $b_1=\beta J_s$ under Eq.~\eqref{eq:elastic-metrics} for $\beta=(k_BT)^{-1}$, and estimates the six coefficients from molecular dynamics trajectories. The second harmonic already allows multiple torsional peaks within one elastic state. Discrete elastic-rod descriptions distinguish centerline geometry from a material orientation around the tangent~\cite{bergou2008discrete}. In a four-$\mathrm{C}_\alpha$ representation, atomic orientation around the central virtual bond is absorbed into the effective harmonic coefficients unless an additional orientation variable is introduced. Setting $a_1=b_1=0$ gives a curvature-independent control, whereas replacing $s=r$ with $x_L+x_R$ or $\min(x_L,x_R)$ tests alternative bend couplings.

Product and sum inputs are related by
\begin{equation}
\label{eq:product-sum}
 x_Lx_R=\bar x(x_L+x_R)-\bar x^2+(x_L-\bar x)(x_R-\bar x).
\end{equation}
Near $(\bar x,\bar x)$, the affine part is absorbed into the coefficients of the sum model, leaving a quadratic difference in the bend deviations that accounts for their similar predictions on folded proteins. Because torsion is defined only when both bends are nonzero, the product $r=x_Lx_R$ makes the bend-dependent interaction vanish continuously as either bend straightens, whereas $x_L+x_R$ remains finite at that boundary even though the background harmonics still require a defined torsion.

A unit is geometrically helical when its minimum root mean square deviation from a right-handed four-C$_\alpha$ template is at most 0.5~\AA\ after translation and proper rotation~\cite{kabsch1976solution,kabsch1978discussion}. The template has radius 2.3~\AA, rise 1.5~\AA\ per residue and rotation $+100^\circ$ per residue. Keeping observed bond lengths and bends fixed and replacing torsion by $t$ gives the indicator $H_i(t)$ and predicted occupancy
\begin{equation}
\label{eq:helix-probability}
 \widehat h_i=\int_{-\pi}^{\pi}H_i(t)p(t\mid s_i)\,dt.
\end{equation}
An isolated gap has predicted probability $h_{i,L}h_{i,R}(1-\widehat h_i)$ conditional on observed neighboring helix labels $h_{i,L},h_{i,R}\in\{0,1\}$.

\subsection{Anisotropic bending and conformational states}
\label{sec:elastic-upgrade}

A single bend response can allocate probability between torsion signs while giving an inaccurate width within either population. We allow different stiffnesses along two directions of the shared normal plane, replacing the scalar stiffness in Eq.~\eqref{eq:elastic-metrics} by a symmetric matrix. With $q=(x_L,x_R)$ and $g_6$ denoting the exponent in Eq.~\eqref{eq:angular-density}, the eight-coefficient angular potential is
\begin{equation}
\label{eq:anisotropic-g8}
 g_8(t,q)=g_6(t,r)+x_R^2(c_2\cos2t+d_2\sin2t),
\end{equation}
where inverse temperature is absorbed into the positive-semidefinite normal-plane stiffness matrix
\begin{equation}
\label{eq:anisotropic-matrix}
 \mathsf K=\begin{pmatrix}a_1&-2d_2\\-2d_2&a_1+4c_2\end{pmatrix}\succeq0.
\end{equation}
Its eigenvalues are $a_1+2c_2\pm2\sqrt{c_2^2+d_2^2}$, so a nonnegative mismatch energy requires $a_1+2c_2\geq2\sqrt{c_2^2+d_2^2}$. In coordinates $\mathbf a_L=(x_L,0)$ and $\mathbf a_R=x_R(\cos t,\sin t)$, defining the torsion-independent bend energy $C(q)=\tfrac12a_1(x_L^2+x_R^2)+c_2x_R^2$ gives the quadratic mismatch expansion
\begin{equation}
\label{eq:anisotropic-expansion}
\begin{aligned}
 \tfrac12(\mathbf a_R-\mathbf a_L)^{\mathsf T}\mathsf K(\mathbf a_R-\mathbf a_L)
 ={}&C(q)-a_1r\cos t+2d_2r\sin t\\
 &-x_R^2(c_2\cos2t+d_2\sin2t).
\end{aligned}
\end{equation}
Subtracting the directed pairing $(b_1+2d_2)r\sin t$ and the angular background $g_0(t)=a_0\cos t+b_0\sin t+c_0\cos2t+d_0\sin2t$ gives $C(q)-g_8(t,q)$. Directional stiffness therefore changes the second harmonic through the right bend as well as restricting its coefficients. This anisotropic response is retained in each component of the two-state elastic model.

The two states represent backgrounds with different preferred bends and elastic responses. Their relative bend energy is $U_-(x)=0$ and $U_+(x)=\delta_0+\delta_1x+\delta_2x^2$; any common bend potential cancels at fixed $q$. For $u,v\in\{-1,+1\}$, independent state priors $w_u=e^{hu}/(2\cosh h)$ and $w_v=e^{hv}/(2\cosh h)$ are set by a global field $h$, giving the unnormalized bend weight for an adjacent state pair
\begin{equation}
\label{eq:local-state-elastic-weights}
 \xi_{uv}(q)=w_uw_v e^{-C_v(q)-U_u(x_L)-U_v(x_R)}.
\end{equation}
Integrating each state potential over torsion defines the state angular partition function $Z_v(q)=\int_{-\pi}^{\pi}e^{g_{8v}(t,q)}dt$, and summing over the adjacent states yields the conditional torsion density
\begin{equation}
\label{eq:state-elastic-density}
 p(t\mid q)=\frac{\sum_{uv}\xi_{uv}(q)e^{g_{8v}(t,q)}}
 {\sum_{uv}\xi_{uv}(q)Z_v(q)}.
\end{equation}
The probability of a state pair at the observed bends is proportional to $\xi_{uv}(q)Z_v(q)$. A favorable bend energy and a larger angular partition function both increase its contribution. The constants $C_v(q)$ cancel within an isolated component but contribute to the relative weights when comparing different components. The two-state elastic model thus changes which elastic background contributes at each bend pair, including the width and location of its torsion peaks.

\section{Results}
\label{sec:results}

\subsection{Baseline local elasticity and torsional order}
\label{sec:atlas}
\label{sec:prediction-error}

The analyzed ATLAS collection contains 1938 protein chains with three 100 ns trajectories per chain~\cite{vander2024atlas}. A one-to-nine fitting-to-test partition keeps sequence-related chains, common-PDB sources and all repeats together, giving 194 fitting chains and 1744 test chains. The test set contains 1317 sequence groups. Predictions are averaged equally over chains within a group and then over groups, with 95\% intervals from 2000 paired group bootstrap resamples. All elastic models use the same fitting observations and are evaluated on the same test observations.

Relative to the curvature-independent control, including the curvature-product interaction reduces the continuous negative log-density loss from 1.490 to 1.160 nats, the mean absolute chain error in negative-torsion fraction from 12.26 to 2.15 percentage points, and geometric helix-fraction error from 21.84 to 10.41 points. These reductions indicate that local bending carries substantial information about torsion direction and helical geometry. Replacing the product by the sum or minimum of the two bends yields smaller changes than introducing bend dependence itself (Table~\ref{tab:atlas-models}).

The remaining structural error depends on how probability is distributed within each torsion direction. For an angular region $I_\sigma$, let $w_\sigma(r)$ be its probability and $q_\sigma(t\mid r)$ its normalized density. Splitting the circle into negative and nonnegative torsion gives the exact identity
\begin{equation}
\label{eq:loss-decomposition}
 -\log p(\tau\mid r)=-\log w_\sigma(r)-\log q_\sigma(\tau\mid r).
\end{equation}
The first term measures allocation between directions and the second measures the angular shape within the observed direction. Circular concentration within a region is $R=|\mathbb E[e^{i\tau}\mid\tau\in I_\sigma]|$. Two models can assign the same directional probability while assigning different probabilities to a narrow helical range.

\begin{figure*}[tp]
\centering
\includegraphics[width=\textwidth]{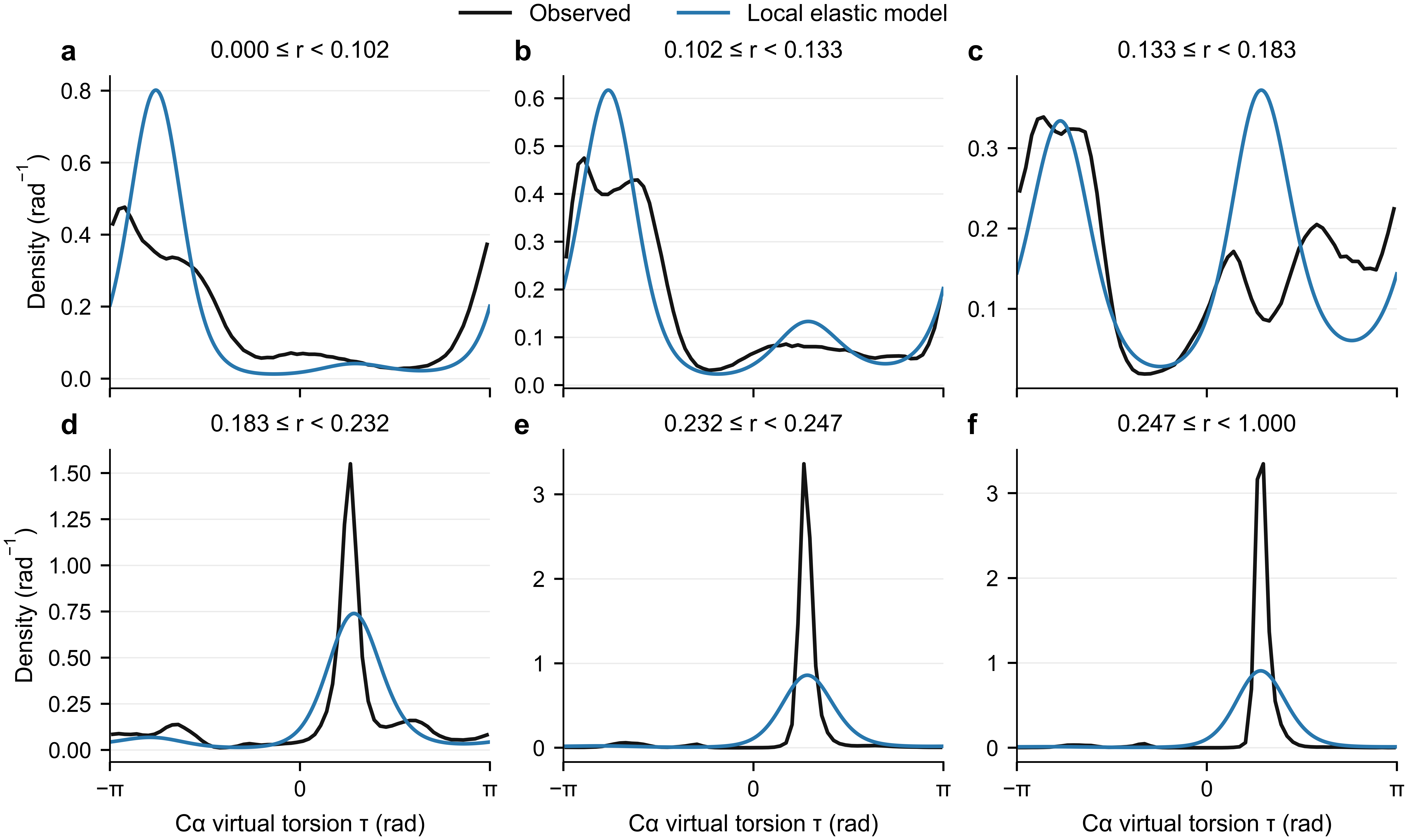}
\caption{Observed and predicted C$_\alpha$ virtual-torsion densities across six ranges of the normalized curvature product $r=x_Lx_R$. The ranges are defined from the fitting data, and predictions use the single-state elastic model. Each chain is normalized within a range before averaging equally over chains within sequence groups and over contributing groups. Observations within each chain and range have equal weight. Low and intermediate ranges show differences in direction allocation. At high curvature, the predicted positive-torsion peaks are broader than the observed peaks. Densities are per radian.}
\label{fig:atlas-distributions}
\end{figure*}

Figure~\ref{fig:atlas-distributions} resolves this difference across six fitting-defined curvature-product ranges. In the lowest range, the single-state elastic model predicts 86.64\% negative torsion against 74.34\% observed. In the intermediate range $0.13302\leq r<0.18268$, these fractions are 44.73\% and 51.90\%. In the two highest ranges, directional errors are about 0.68 percentage points, yet observed geometric helicities of 90.64\% and 94.09\% are predicted as 66.36\% and 69.01\%. The broader predicted positive-torsion peaks account for the appreciable helix error that remains after directional allocation improves.

\subsection{Two-state elastic predictions and reference backbone models}
\label{sec:external-models}
\label{sec:upgrade-results}

\begin{figure*}[tp]
\centering
\includegraphics[width=.92\textwidth]{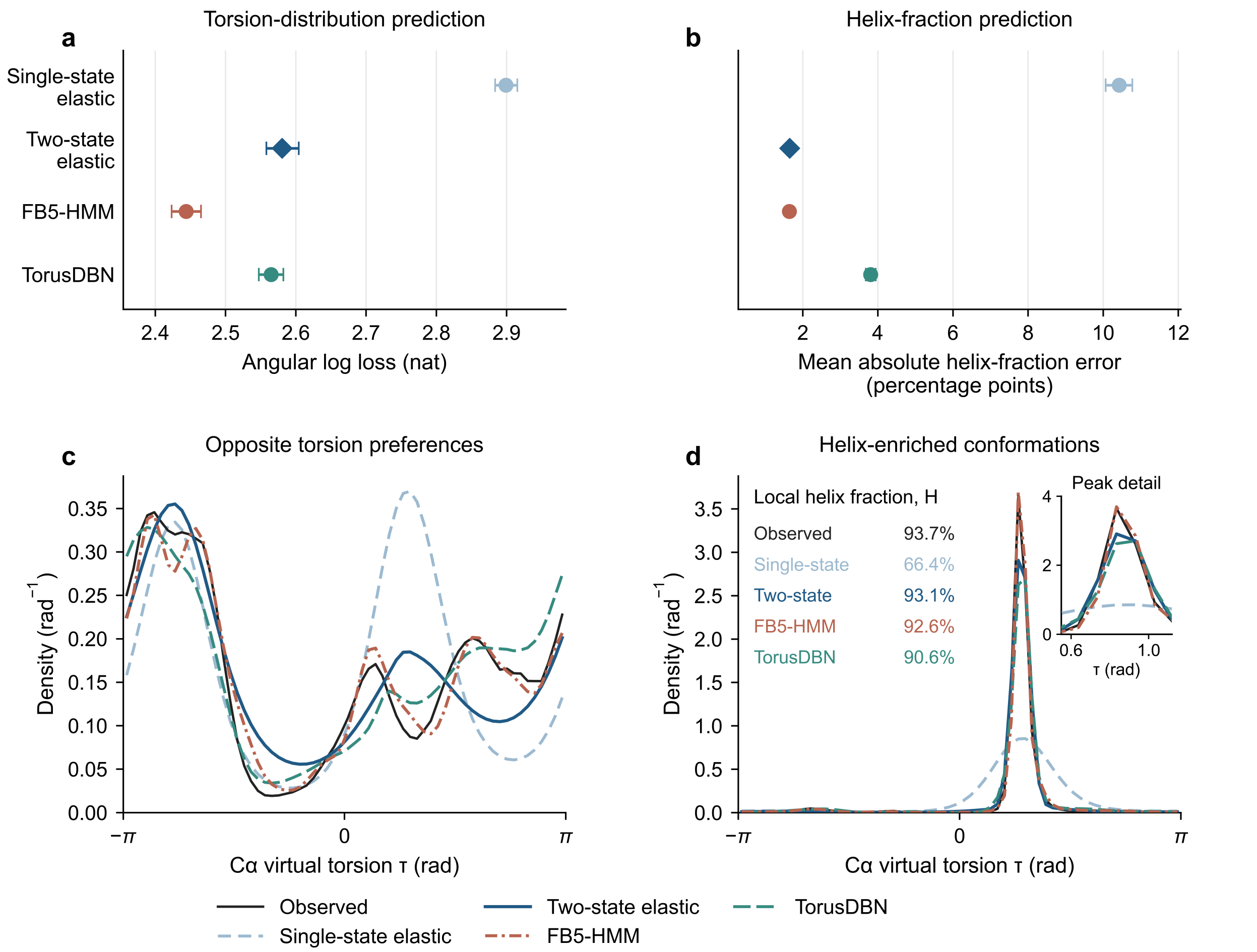}
\caption{Two-state elastic predictions and reference backbone models on 1744 ATLAS test chains. Panel a shows negative log probability in 36 torsion intervals and panel b shows absolute chain geometric helix-fraction error. All models use the same 100-cell bending support, with equal chains within sequence groups and equal groups. Intervals are 95\% limits from 2000 paired sequence-group bootstrap resamples. Panels c and d show distributions at $0.13302\leq r<0.18268$ and $0.23214\leq r<0.24694$, respectively, using native-coordinate range assignment and common within-range support. Curves join the centers of 64 angular bins without fitted smoothing. The inset in panel d enlarges the positive-torsion peak using the same bin values. Range conditioning follows cohort averaging, so each chain contributes in proportion to its frequency within the range, unlike the chain-normalized densities in Figure~\ref{fig:atlas-distributions}. The retained observed mass within the two ranges is 99.98\% and 99.99\%. Densities are per radian.}
\label{fig:upgrade-validation}
\end{figure*}

The two-state elastic model allows the observed bend pair to select between different bend preferences and anisotropic responses through Eq.~\eqref{eq:state-elastic-density}. Figure~\ref{fig:upgrade-validation} compares it with the single-state elastic model and the published FB5-HMM and TorusDBN references. All four models use a common grid of 100 bending cells and 36 torsion intervals on the 1744 test chains, with mean common bending coverage of 99.81\% and a minimum of 97.47\% for an individual chain. Angular interval loss is 2.899 nats for the single-state elastic model, 2.581 for the two-state elastic model, 2.444 for FB5-HMM and 2.565 for TorusDBN. The corresponding mean absolute chain helix-fraction errors, obtained by summing probabilities within each chain before comparison, are 10.43, 1.65, 1.64 and 3.81 percentage points (Table~\ref{tab:common-four-models}).

Panels c and d of Figure~\ref{fig:upgrade-validation} resolve this helical improvement across intermediate and high curvature-product ranges assigned from the native coordinates of each sample. All curves use the same observed bending weights and common support. In the high range, observed helicity is 93.74\%, compared with 66.44\% for the single-state elastic model, 93.10\% for the two-state elastic model, 92.57\% for FB5-HMM and 90.62\% for TorusDBN. The two-state elastic model places substantially more probability mass in the narrow helical region while preserving a normalized angular distribution.

Direct continuous evaluation across all observed targets without a bending-support restriction is consistent with this pattern (Table~\ref{tab:elastic-upgrades}). Extending the single-state model to two states lowers continuous density loss from 1.160 to 0.790 nats, reduces mean absolute chain helix-fraction error from 10.41 to 1.65 percentage points, and lowers the unit-level helix Brier score, the mean squared error of predicted occupancy, from 0.0708 to 0.0432. The two-state elastic model thus approaches the 1.56-point chain helix error and the 0.0366 Brier score of published FB5-HMM while retaining a larger continuous angular loss than its 0.448 nats.

At fixed bend, each Fisher--Bingham component of FB5-HMM has the same first two log-density harmonics as the elastic density, whereas its bend response and mixture weights vary with hidden state and sequence. Frozen-reference ablations indicate that neighboring-state dependence and sequence information each contribute to this difference, as removing adjacent-state dependence and marginalizing amino-acid information raise the fitting-set FB5-HMM loss of 0.430 nats by 0.283 and 0.167 nats, respectively. Circular-moment projections further indicate that restricting the coefficients to the global six-coefficient form accounts for a larger divergence than angular structure outside the two-harmonic family. Single-state anisotropic controls and a sequence-associated two-state variant separate the effects of directional stiffness, the positive-stiffness constraint and sequence-dependent state priors (Table~\ref{tab:elastic-upgrades}). These comparisons support distinct elastic backgrounds and directional stiffness without identifying a fitted state with a unique atomic structure.

Standard Lifson--Roig theory addresses residue-level helical populations and cooperativity using its atomic $\phi/\psi$ state definition, which differs from the four-$\mathrm{C}_\alpha$ geometric criterion used for the angular models. Under that same definition for prediction and observation, it gives a mean helical fraction of 41.48\% against 41.22\% observed (Table~\ref{tab:lr-main}). This agreement in the ensemble mean coexists with an 11.23-point mean absolute chain-fraction error and a 34.01-point residue occupancy error. The mean chain run-length total variation, half the summed absolute probability difference, is 0.275. The comparison distinguishes agreement in total helical population from agreement in its distribution among chains and along a sequence (Table~\ref{tab:standard-lr}).

\begin{table}[tp]
\centering\small
\caption{Standard Lifson--Roig predictions using identical atomic $\phi/\psi$ residue states for model and observation. Errors are in percentage points. Intervals are 95\% sequence-group bootstrap limits.}
\label{tab:lr-main}
\begin{tabularx}{\columnwidth}{@{}Xl@{}}
\toprule
Observable & Estimate\\
\midrule
Mean helix fraction, predicted / observed & 41.48\% / 41.22\%\\
Absolute chain-fraction error & 11.23 (10.77, 11.67)\\
Residue occupancy error & 34.01 (33.74, 34.27)\\
Chain run-length total variation & 0.275 (0.270, 0.281)\\
\bottomrule
\end{tabularx}
\end{table}

\subsection{Transfer across systems, starting conformations and simulation protocols}
\label{sec:simulation}
\label{sec:comparison}

We test transfer from ATLAS native-state ensembles to isolated peptides and a folded miniprotein using fixed ATLAS parameters. The systems are the 21-residue alanine-rich Fs peptide, the 15-residue peptide $(\mathrm{AAQAA})_3$~\cite{shalongo1994distribution}, and the 35-residue villin headpiece HP35 with a proline-containing turn and a packed three-helix structure~\cite{chiu2005high}. Each system has four starting conformations (Table~\ref{tab:data}) and three independent repeats under each of two protocols, giving 72 trajectories of 500 ns and 36 $\mu$s in total. The controlled protocols use AMBER99SB-ILDN/TIP3P and CHARMM36m with matched water, whereas ATLAS uses CHARMM36m/TIP3P.

\begin{figure*}[!t]
\centering
\includegraphics[width=.96\textwidth]{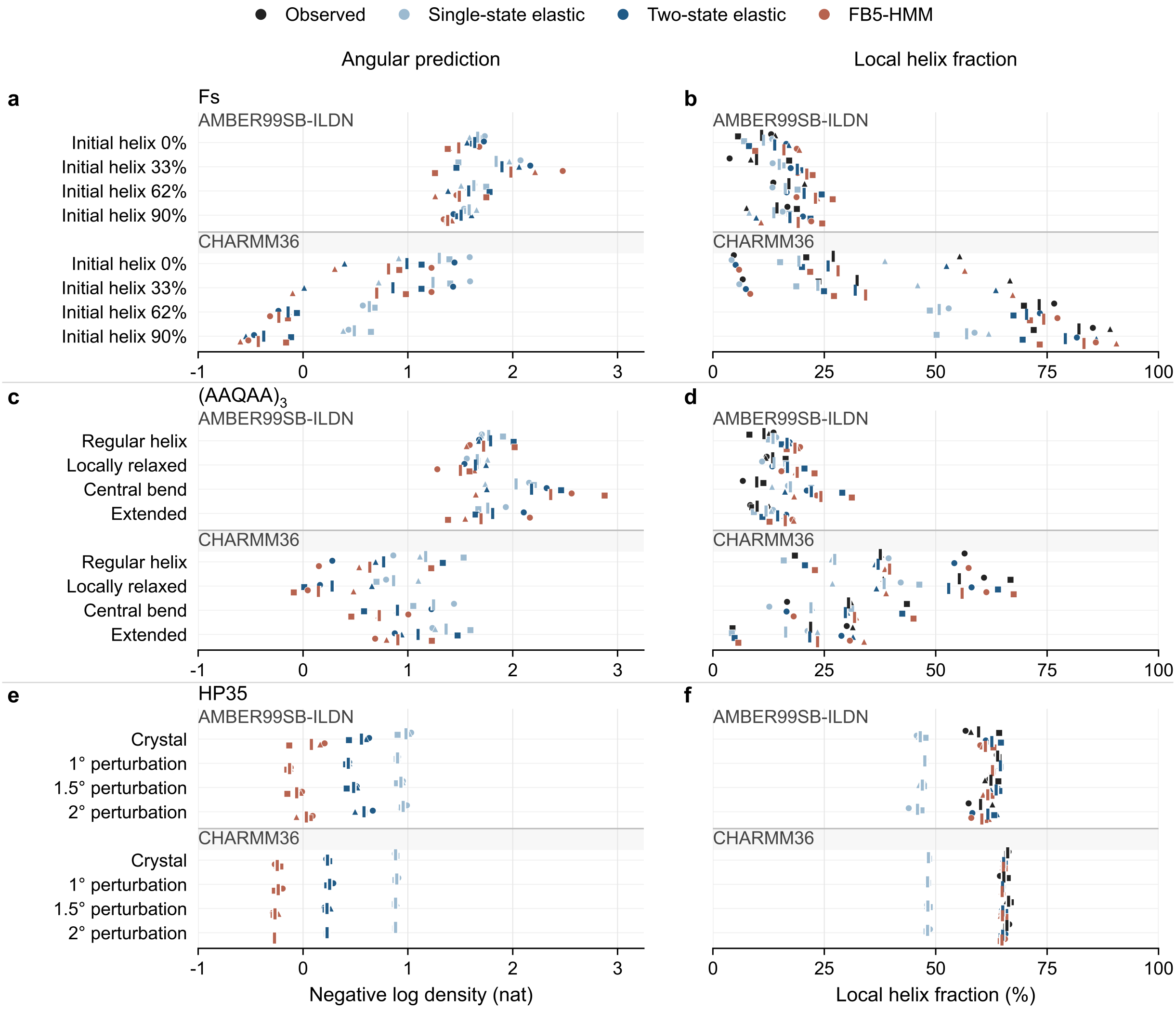}
\caption{Transfer of fixed ATLAS parameters across systems, starting conformations and simulation protocols. Rows show Fs, $(\mathrm{AAQAA})_3$ and HP35. The left column, \emph{Angular prediction}, compares continuous negative log-density loss for the single-state elastic model, two-state elastic model and published FB5-HMM. The right column, \emph{Local helix fraction}, compares observed geometric helix fractions with the same three predictions. Each protocol contains four starts with three repeats per start. Different symbols identify the three repeats, and vertical bars mark the mean for each model within each start and protocol group. Both columns use the same 72 trajectories, all internal four-C$_\alpha$ units and every tenth saved frame at 0.1 ns intervals over $5<t\leq500$ ns. The controlled protocols are AMBER99SB-ILDN/TIP3P and CHARMM36m with matched water.}
\label{fig:controlled}
\end{figure*}

Figure~\ref{fig:controlled} evaluates the fixed ATLAS parameters without refitting on all internal four-$\mathrm{C}_\alpha$ units of Fs, $(\mathrm{AAQAA})_3$ and HP35 at 0.1 ns intervals over $5<t\leq500$ ns. Averaged equally over the 24 system, protocol and start combinations, the two-state extension reduces continuous density loss from 1.231 to 0.892 nats and mean absolute trajectory helix-fraction error from 10.73 to 2.87 percentage points, compared with 0.658 nats and 3.34 points for published FB5-HMM (Table~\ref{tab:elastic-upgrades}). The lower FB5-HMM angular loss indicates that further conditional angular structure remains unresolved. Across the 72 individual trajectories, the two-state elastic model lowers continuous angular loss in 64 and trajectory helix-fraction error in 48, suggesting that angular-density gains transfer more consistently than gains in total local helicity.

Comparing the two force-field protocols indicates that transfer accuracy depends on the sampled conformational regime. Observed Fs helicity changes from 13.03\% under AMBER to 53.67\% under CHARMM, compared with two-state elastic predictions of 17.63\% and 51.81\%, respectively. In the coil-rich AMBER ensembles, replacing the single-state elastic model with the two-state elastic model increases mean absolute trajectory helix-fraction error from 2.49 to 4.63 percentage points for Fs and from 3.14 to 6.33 points for $(\mathrm{AAQAA})_3$. Under CHARMM, the two-state elastic model improves helix-fraction predictions for all three systems (Figure~\ref{fig:controlled}). The excess predicted helicity in the two AMBER peptide ensembles points to a protocol-dependent bias in the fixed two-state elastic model, even when the aggregate angular score improves.

The controlled trajectories retain differences associated with their starting structures and simulation protocols. The comparison therefore tests conditional predictions in the sampled ensembles rather than an equilibrium folding population (Tables~\ref{tab:controlled-external} and \ref{tab:controlled-lr}).

\subsection{Atomic context at matched bends and material orientation}
\label{sec:atomic-context}

To examine the structural information remaining at similar bends, we use Fs residues 16 to 19, $(\mathrm{AAQAA})_3$ residues 12 to 15 and HP35 residues 20 to 23. The latter two windows maximize shared two-bend coverage across the eight start and protocol combinations. Negative-torsion peptide frames are matched to their nearest nonnegative controls using both bends. At tolerance 0.002, each bend differs by at most $0.36^\circ$, yet median circular torsion separations range from $92.7^\circ$ to $118.2^\circ$. Negative populations have circular concentration $R=0.795$ to 0.853 and matched controls have $R=0.470$ to 0.537. Similar bending therefore permits distinct, internally ordered torsional populations.

\begin{figure*}[!t]
\centering
\includegraphics[width=.96\textwidth]{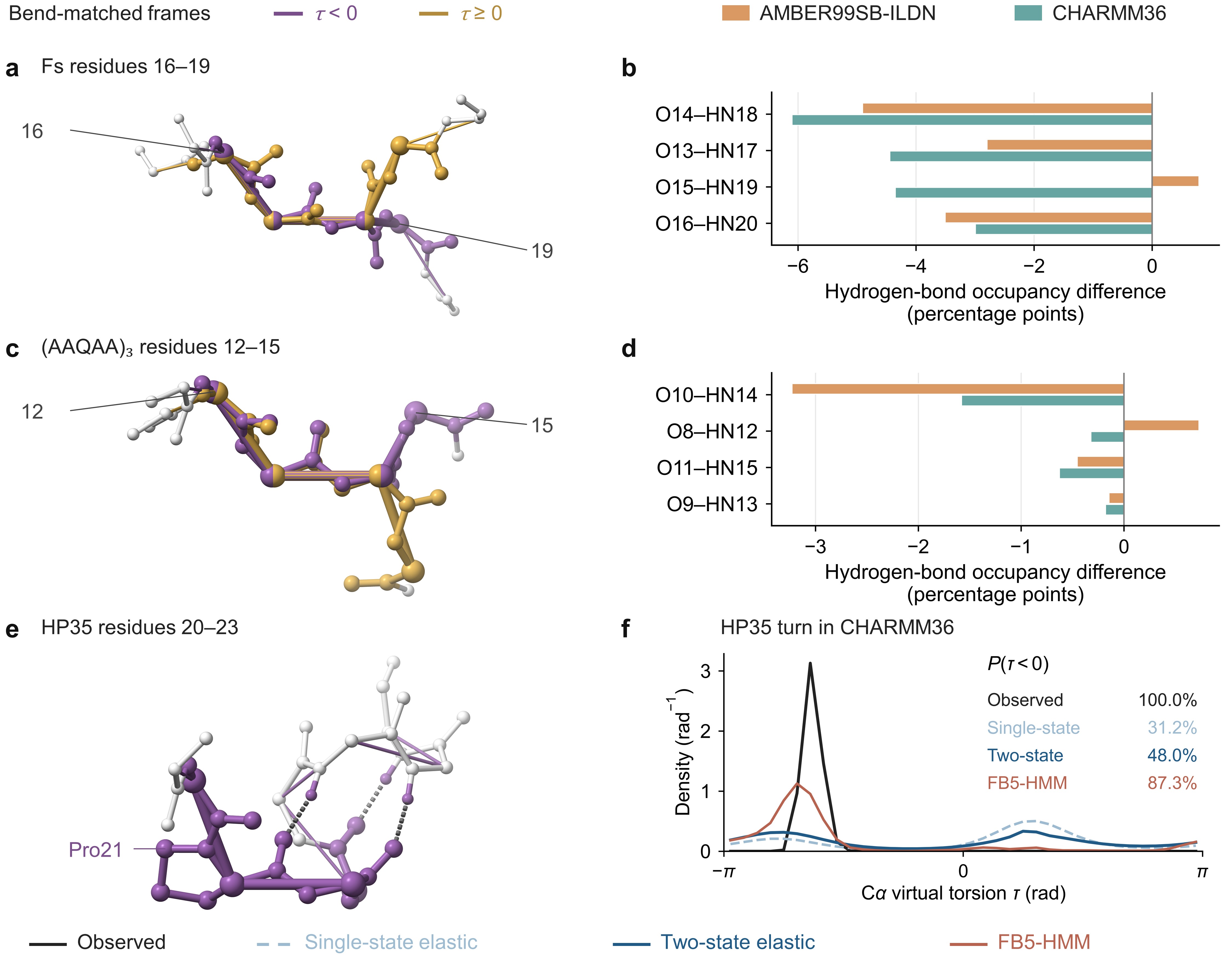}
\caption{Atomic context at matched bends and the ordered HP35 turn. Panels a and c superpose actual bend-matched CHARMM frames of Fs residues 16--19 and $(\mathrm{AAQAA})_3$ residues 12--15 rendered in ChimeraX. Purple and warm gold identify negative and nonnegative torsions, respectively; neighboring backbone atoms are gray. Panels b and d give backbone hydrogen-bond occupancy differences between negative frames and matched controls in percentage points. The difference is negative minus nonnegative, and O$_i$--HN$_j$ identifies the acceptor and donor. Warm orange denotes AMBER99SB-ILDN and blue-green denotes CHARMM36m. Panel e enlarges the Pro21 environment in the HP35 20--23 turn; gray dashed bonds satisfy the hydrogen-bond criterion in the displayed frame. Panel f compares the same CHARMM HP35 unit using observed density in black, the single-state elastic model in light blue, the two-state elastic model in dark blue and published FB5-HMM in brick red. Figure~\ref{fig:hp35-amber} gives the corresponding AMBER distributions.}
\label{fig:matched-atomic}
\end{figure*}

The matched CHARMM examples in Figure~\ref{fig:matched-atomic}a and c show the corresponding atomic arrangements. Circular means of $\phi_{18}$ differ by $38^\circ$ to $40^\circ$ in Fs and $\phi_{14}$ by $48^\circ$ to $64^\circ$ in $(\mathrm{AAQAA})_3$ across the two protocols. Backbone hydrogen-bond occupancies also differ. Fs O14--HN18 occupancy is 4.90 percentage points lower in negative frames under AMBER and 6.10 points lower under CHARMM than in their matched controls. Panels b and d resolve these contact-specific differences, indicating that opposite torsional populations at matched bends are associated with distinct backbone hydrogen-bond patterns (Table~\ref{tab:controlled-matching}).

The HP35 window contains Pro21 and forms an ordered nonhelical turn that remains at negative torsion in all 594\,000 densely saved frames per protocol, with $R\simeq0.993$ and zero geometric helicity. On the common 0.1 ns target, the single-state elastic model assigns 33.00\% negative probability under AMBER and 31.16\% under CHARMM. The two-state elastic model increases these probabilities to 49.35\% and 48.02\%, whereas direct FB5-HMM gives 87.67\% and 87.26\% (Table~\ref{tab:selected-turns}). The state mixture reduces a local directional error but leaves a substantial discrepancy. Figure~\ref{fig:matched-atomic}e shows the atomic environment of this turn and panel f resolves its CHARMM angular density. The proline-containing turn and surrounding contacts suggest that this unresolved directional preference reflects specific atomic constraints.

A projected C$_\alpha$--N direction provides one possible additional local variable. Its material angle $d$ measures the nitrogen direction in the second residue around the shared virtual bond, relative to $\mathbf n_L=\mathbf b_L\times\mathbf t$. For fixed bends $q=(x_L,x_R)$, a bilinear coupling to the right normal $\mathbf n_R=\mathbf b_R\times\mathbf t$ gives the coupled energy
\begin{equation}
\label{eq:material-energy}
 E(q,\tau,d)=E_0(q,\tau)+E_m(q,d)-Jr\cos(\tau-d),
\end{equation}
where the baseline $E_0(q,\tau)$ corresponds to the single-state elastic density $p_0(\tau\mid q)=Z_0(q)^{-1}e^{-\beta E_0(q,\tau)}$ in Eq.~\eqref{eq:angular-density}. When the material orientation is frozen at a common angle $d_*$, the coupling reparameterizes the existing curvature slopes according to
\begin{equation}
\label{eq:material-energy-frozen}
 (a_1^{\mathrm{eff}},b_1^{\mathrm{eff}})=(a_1,b_1)+\beta J(\cos d_*,\sin d_*).
\end{equation}
In the controlled trajectories, the material-angle concentration $R_d=|\langle e^{id}\rangle|$ ranges from 0.986 to 0.997. An empirical $L^2$ Gram projection places 96.7\% to 99.6\% of the added coefficient-function squared norm within the original six-function span. The projection measures redundancy on these sampled geometries, consistent with the reparameterization at a fixed material angle.

For a fluctuating direction, define $w_0(d\mid q)\propto e^{-\beta E_m(q,d)}$, $z=\beta Jr$, $m_n=\mathbb E_{w_0}e^{ind}$ and $\Lambda=\mathbb E_{w_0}e^{z\cos(\tau-d)}$. Then $p\propto p_0\Lambda$, with
\begin{equation}
\label{eq:material-cumulants}
\begin{aligned}
 \log\Lambda={}&z\operatorname{Re}(m_1^*e^{i\tau})\\
 &+\frac{z^2}{4}\{1-|m_1|^2\\
 &\quad+\operatorname{Re}[(m_2^*-(m_1^*)^2)e^{2i\tau}]\}+O(|z|^3).
\end{aligned}
\end{equation}
For this coupling, a fixed material direction shifts the existing first-harmonic coefficients. Fluctuations generate a second-harmonic correction whose leading amplitude scales as $r^2$ when the material moments remain fixed across bends. The tested anisotropic mismatch contributes a term proportional to $x_R^2$. The material reduction thus specifies a distinct possible correction. Its magnitude is not determined by the observed material-angle concentration alone because the fluctuation term depends on the uncoupled moments $m_1$ and $m_2$, which satisfy $|m_2-m_1^2|\leq1-|m_1|^2$ and therefore suppress the correction only as the uncoupled distribution becomes fully concentrated.

\section{Discussion}
\label{sec:discussion}

Formulating discrete Hasimoto frame transport as an elastic Hamiltonian bridges the geometric description of static protein backbones with the statistical mechanics of fluctuating ensembles. Whereas discrete nonlinear Schr\"odinger models describe individual conformations as deterministic discrete solitons, evaluating the rotational mismatch between neighboring bending generators in $\mathfrak{so}(3)$ directly yields an analytical statistical potential for conformational trajectories. While curvature--torsion order parameters and local entropy capture overall conformational diversity~\cite{grassein2020curvature,senet2026local}, this conditional formulation resolves a further distinction between torsion-direction probability and angular concentration within that direction. On ATLAS, the single-state elastic baseline improves direction allocation while leaving helical peaks too broad, and its residual helix-fraction error indicates that improved direction probabilities can still leave structural populations inaccurately described.

Different bend preferences and anisotropic elastic responses improve prediction when their contributions are weighted by both bend energy and the angular partition function through $\xi_{uv}(q)Z_v(q)$. A state with lower bend energy need not dominate if the competing state has greater integrated angular weight. This dependence of effective interactions on conformational background is also central to statistical-mechanical treatments of multiple coarse-grained free-energy surfaces~\cite{jin2023statistical}. In the two-state elastic model, the two states supply different elastic responses to observed bends, while each state can already support multiple torsional peaks. The close helix-fraction accuracy of the two-state elastic model and FB5-HMM, together with their different angular losses, indicates that matching one structural population can leave further angular information unresolved.

Helical population accuracy also leaves correlations between neighboring backbone units unresolved. Coarse-graining studies of peptide ensembles distinguish local structural agreement from helix--coil cooperativity~\cite{rudzinski2015bottom}, and multitorsional UNRES potentials improve modeled helix and linker geometry through interactions between successive torsions~\cite{petrusevich2025recapturing}. The reference models represent these correlations at different resolutions. FB5-HMM uses sequence and neighboring hidden states to vary continuous angular responses~\cite{hamelryck2006sampling}. TorusDBN predicts atomic backbone variables before conversion to C$_\alpha$ geometry~\cite{boomsma2008generative}. Lifson--Roig theory describes residue-level helix--coil cooperativity~\cite{lifson1961theory,vitalis201250}; in the present comparison it reproduces the mean residue-state helical population, with remaining differences in its allocation among chains, residue positions and run lengths.

Across the controlled trajectories, the two-state elastic model lowers the aggregate angular loss and mean absolute helix-fraction error relative to the single-state elastic model using fixed ATLAS parameters, whereas the helix-fraction error increases for both peptides under AMBER. Peptide force-field benchmarks report that conformational populations and folding behavior depend on the force field, sequence and starting structure~\cite{best2009optimized,singh2026well}, consistent with this uneven transfer of the fixed interaction across the sampled conformational ensembles. Start-dependent populations restrict this interpretation to conditional local predictions in those ensembles, without determining equilibrium folding free energies or transition rates.

The atomic comparisons identify structural differences at similar bends. Fs and $(\mathrm{AAQAA})_3$ retain opposite torsion populations with different atomic dihedrals and hydrogen-bond occupancies. Probabilistic C$_\alpha$ backmapping samples possible atomic structures from reduced coordinates~\cite{jones2023diamondback}, and recent reconstruction benchmarks locate deviations in proline geometry and cooperative secondary structure when atomic ensembles are recovered from single-bead simulations~\cite{huang2026local}. Consistent with the geometric restrictions imposed by proline on the polypeptide backbone~\cite{macarthur1991influence}, the negative-torsion HP35 turn remains underestimated after the two-state extension. Static-structure analyses also associate torsion-sign changes between adjacent positions with curvature variation and nonlocal contacts~\cite{wang2026instability}, motivating an examination of neighboring geometry and contact environment alongside the local bends. Separating the contributions of Pro21, hydrogen bonding and tertiary packing in HP35 would require controlled comparisons across sequence variants or contact environments at matched geometry.

Eliminating the material angle specifies how an additional atomic orientation can enter the local potential. Averaging peptide-group orientations has previously yielded multitorsional interactions through cumulant expansions~\cite{sikorska2022origin}. In the present reduction, a common fixed direction shifts the existing first-harmonic slopes, and fluctuations introduce a second-harmonic term that scales as $r^2$ when the uncoupled moments are bend-independent. An empirical Gram projection onto the original six-function span recovers most of the squared $L^2$ norm of the added orientation-dependent coefficient functions. The fixed-angle identity supplies a possible explanation for this redundancy. The concentration measured from the coupled trajectories does not determine the material moments of the uncoupled distribution, including $m_1$, which are needed to evaluate the fluctuation correction. The predictive value of the additional variable depends on the information and couplings it contributes beyond the retained bend-dependent angular functions, connecting this example to the broader question of representability in coarse-graining~\cite{noid2024rigorous}. More broadly, transforming a discrete geometric identity into an analytical statistical potential demonstrates how exact differential-geometric relations can be generalized to describe conformational thermodynamics in fluctuating biomolecules.

\Needspace{12\baselineskip}
\section{Methods}
\label{sec:methods}

\subsection{Dataset partition and molecular dynamics protocols}
\label{app:simulation}
\label{app:controlled-protocol}

The ATLAS comparison uses sequence groups formed from MMseqs2 relationships at 30\% identity, 80\% bidirectional coverage and an E value at most 0.001, together with identical sequences and common PDB sources. All three molecular dynamics repeats of a chain remain in the same partition. The fit contains 194 chains in 139 sequence groups, and evaluation contains 1744 chains in 1317 groups. All elastic forms, including the extensions, are fitted to the same 9894 observations comprising 51 equally weighted units dispersed across positions, repeats and times after 5 ns from each fitting chain. Each test chain supplies 4000 saved evaluation observations restricted to times after 5 ns. Model forms and numerical settings were chosen in an earlier grouped exploration with 1165 fitting, 388 selection and 385 evaluation chains; the present comparison refits these fixed forms on the 194-chain subset and reports all evaluated elastic variants. Full-model adaptation of external generative references uses twelve complete atomic conformations per fitting chain, with four times from each repeat after 5 ns, assigning equal total weight to each sequence group, equal weight to chains within each group and equal weight to sampled conformations within each chain, divided by chain length.

For the controlled simulations, Fs contains 21 residues with sequence \texttt{AAAAAAAARAAAARAAAARAA} and ACE/NME caps, $(\mathrm{AAQAA})_3$ contains 15 residues with ACE and amide caps, and HP35 is the 35-residue N27H variant derived from PDB ID 1YRF with free termini and neutral epsilon-protonated histidine~\cite{chiu2005high}. Table~\ref{tab:data} summarizes the four starting conformations per system. All three repeats of a fixed starting structure use separately recorded solvent, velocity and thermostat random states; the first two Fs repeats per start and condition retain their existing 40 ns production and native random states when available. The Fs starts use the conformations at 249.95 ns from archive trajectories 5, 21, 27 and 1, respectively~\cite{fsarchive}. The regular $(\mathrm{AAQAA})_3$ helix has $(\phi,\psi)=(-57^\circ,-47^\circ)$ and trans peptide bonds; the locally relaxed start substitutes $(-75^\circ,-25^\circ)$ at residues 4, 5, 11 and 12; the central-bend start substitutes $(-80^\circ,-20^\circ)$, $(-100^\circ,10^\circ)$ and $(-75^\circ,-30^\circ)$ at residues 7, 8 and 9; and the extended start uses $(-135^\circ,135^\circ)$ throughout. HP35 start 1 retains the experimental heavy atoms with the missing terminal oxygen added. For starts 2 to 4, indexed by $j=1,2,3$ with amplitudes $(A_1,A_2,A_3)=(1^\circ,1.5^\circ,2^\circ)$, backbone increments at residue $i$ are
\begin{equation}
\label{eq:hp-starts}
 \delta\theta_{ijc}=A_j\sin(\varphi i+0.73j+c),
\end{equation}
where $\varphi=(1+\sqrt{5})/2$, $c=0$ for $\phi$ and $c=1.1$ for $\psi$. Undefined terminal torsions and proline $\phi$ rotations are omitted, and hydrogens are rebuilt with heavy atoms fixed. The two controlled protocols use AMBER99SB-ILDN/TIP3P and CHARMM36m protein parameters~\cite{huang2017charmm36m} from the OpenMM files derived from the July 2024 CHARMM release, with matched CHARMM-modified TIP3P water. Both use 1.2 nm solvent padding and neutralizing ions without added salt. Preparation comprises minimization, 0.1 ns restrained constant-volume equilibration, 0.5 ns restrained constant-pressure equilibration and 1 ns unrestrained constant-pressure equilibration. Production uses OpenMM 8.3.1 at 300 K and 1 bar, a 2 fs step, Langevin-middle friction of 1 ps$^{-1}$, particle-mesh Ewald electrostatics, a 1.2 nm cutoff and a potential switch from 1.0 to 1.2 nm, saving coordinates every 10 ps. ATLAS uses CHARMM36m/TIP3P with 150 mM salt in GROMACS~\cite{vander2024atlas}.

\subsection{Regularized estimation and two-state elastic inference}
\label{app:numerics}
\label{app:elastic-states}

The single-state elastic model and its three curvature controls minimize the mean negative log density plus a coefficient-function penalty with regularizer $\lambda=10^{-4}$. The curvature-independent control sets $a_1=b_1=0$, while the sum and minimum controls replace $s=r$ by $s=x_L+x_R$ and $s=\min(x_L,x_R)$, respectively. With $\bm\eta=(a_0+a_1s,b_0+b_1s,c_0,d_0)$,
\begin{equation}
\label{eq:objective}
 \mathcal L=-\frac1m\sum_i\log p(\tau_i\mid s_i)
 +\frac{\lambda}{4m}\sum_i\lVert\bm\eta(s_i)\rVert^2.
\end{equation}
For $\eta(s)=\mathbf f(s)^{\mathsf T}\bm\beta$ and $G=m^{-1}\sum_i\mathbf f(s_i)\mathbf f(s_i)^{\mathsf T}$, training-only Gram whitening uses $\bm\beta=G^{-1/2}\bm\theta$, so the penalty is $\lambda\lVert\bm\theta\rVert^2/4$ on the nonsingular feature space. Circular fitting and prediction use 512 and 1024 equally spaced midpoint nodes, subtracting the exponent maximum before normalization. Table~\ref{tab:fixed-coefficients} records the fixed parameters, and $\bar x=0.4157$ is the fitting reference in Eq.~\eqref{eq:product-sum}.

The anisotropic energy expansion is given in Eq.~\eqref{eq:anisotropic-expansion}. Its constrained single-state fit is convex in the raw coefficients with the convex matrix constraint in Eq.~\eqref{eq:anisotropic-matrix}. The free anisotropic control uses the identical eight angular coefficients without that constraint. The linear second-harmonic control instead replaces $x_R^2$ by $r$ in Eq.~\eqref{eq:anisotropic-g8}. Each state of the two-state elastic model retains the constrained anisotropic response and parameterizes its stiffness by a Cholesky factor with positive diagonal.

The main two-state elastic model has 20 fitted coefficients, comprising two eight-coefficient angular responses, three relative bend-energy coefficients and one global state field. Its prior factorizes as $w_uw_v$. The optional sequence-associated variant has 40 coefficients and replaces that prior by adjacent-state probabilities from an open-chain Ising distribution,
\begin{equation}
\label{eq:sequence-state-prior}
 P(\bm z\mid A)\propto\exp\!\left(\sum_i h_{A_i}z_i+J_{\rm seq}\sum_i z_i z_{i+1}\right),
\end{equation}
where $A$ is the full amino-acid sequence and $z_i\in\{-1,+1\}$. Twenty amino-acid fields and one neighbor coupling replace the global field. Forward--backward messages give $W_{uv}(A)$ for the central residue pair before its observed bends enter. In this variant,
\begin{equation}
\label{eq:state-elastic-weights}
 \xi_{uv}(q,A)=W_{uv}(A)e^{-C_v(q)-U_u(x_L)-U_v(x_R)},
\end{equation}
which replaces Eq.~\eqref{eq:local-state-elastic-weights} in the density and normalization. Each prediction uses the full sequence and its own bend pair; other observed torsions do not enter it. The local model is obtained with $J_{\rm seq}=0$ and constant fields, giving $W_{uv}=w_uw_v$.

For the two-state model, the dimensionless edge energy is
\begin{equation}
\label{eq:edge-energy}
 E_{uv}(q,t)=C_v(q)+U_u(x_L)+U_v(x_R)-g_{8v}(t,q).
\end{equation}
Summing $w_uw_ve^{-E_{uv}}$ over the states and integrating over $t$ gives Eq.~\eqref{eq:state-elastic-density}, with predictive edge probabilities proportional to $\xi_{uv}Z_v$. Replacing $w_uw_v$ by $W_{uv}(A)$ gives the sequence-associated variant. This reduction retains the state-dependent constants in both $C_v$ and $U_v$; if $g_{8v}$ is common to the states, it factors from the sum and recovers its normalized single-state density. The open-chain Ising recursion begins with $\alpha_1(v)=h_{A_1}v$ and propagates
\begin{equation}
\label{eq:ising-recursion}
 \alpha_{i+1}(v)=h_{A_{i+1}}v+\log\sum_u e^{\alpha_i(u)+J_{\rm seq}uv},
\end{equation}
while the backward recursion includes fields to the right of the queried edge to give $W_{uv}$ before the local bend update. Normalized log messages and their derivatives with respect to the 20 fields and $J_{\rm seq}$ are propagated together and checked against short-chain enumeration and finite differences. In the 20-coefficient local model, the constant field $h$ and relative energy constant $\delta_0$ enter through $2h-\delta_0$, with regularization selecting a representative. The fitting objective adds $\lambda/8$ times the mean squared four-coefficient norm summed over states ($\lambda=10^{-4}$) and $\lambda/4$ times the mean squared centered edge energy, the mean squared fields and $J_{\rm seq}^2$ when present. Coefficients are optimized with bounded limited-memory quasi-Newton optimization (L-BFGS) from separate sign-based and identical-component perturbation starts, restricting sequence fields to $[-4,4]$ and $J_{\rm seq}$ to $[0,3]$ and retaining the lower converged training objective.

\subsection{Common bending support and structural probabilities}
\label{app:common-evaluation}

The comparison in Figure~\ref{fig:upgrade-validation}a and b uses the 100 finer cells obtained by splitting each interval of the fitting-defined $5\times5$ bending grid at its midpoint. A cell is retained when both sampling batches of every published and fitting-set-adapted reference generator contain native-coordinate observations in that cell. The same retained observations are used for all four models. Both elastic models are evaluated at the individual observed bend pairs and integrated using 1024 angular nodes. The external distributions reuse two independent batches of 512 complete conformations per test sequence. Native generated coordinates determine their bends, torsions and geometric helix labels. The 36-interval probabilities use a symmetric pseudocount of one half per interval, equivalent to total uniform prior mass 18. All four predictions are mixed with the same observed fine-cell weights within each chain. Summing the predicted and observed helix contributions before taking their absolute difference gives the chain-fraction error. Averaging this error equally over chains within each group and then over groups gives Figure~\ref{fig:upgrade-validation}b. The cellwise absolute errors in Table~\ref{tab:external-models} instead take the absolute difference before summing over bending cells.

The two displayed ranges in Figure~\ref{fig:upgrade-validation}c and d are $0.13302\leq r<0.18268$ and $0.23214\leq r<0.24694$, with full-precision boundaries retained in the analysis files. Each observed or generated sample is assigned using $r$ from its own native coordinates. For a given range and fine cell, both independent batches of each native method must contain positive sample counts. All observed and predicted curves then retain the same range-specific bending support and observed cell weights. The angular display uses 64 bins with total uniform prior mass 18 for the external estimates, and integrates internal predictions on 1024 nodes. Lines join bin centers without a fitted smoothing function. The supported joint angular mass is first aggregated with equal chain weights within sequence groups and equal group weights, and only then normalized by the cohort mass in the range. This ordering retains the contribution of each chain to that range instead of assigning equal weights to separately normalized regional curves. The retained observed masses are 99.98\% and 99.99\%, respectively. The 36-interval scores, chain errors and bootstrap limits are recorded in Table~\ref{tab:common-four-models}.

For all interval scores, a torsion observation contributes the negative logarithm of its interval probability. Continuous scores use density with respect to $dt$ in radians.

\subsection{External generative models and Lifson--Roig evaluation}

FB5-HMM uses 75 hidden states, Markov transitions and five-parameter Fisher--Bingham distributions on the sphere of virtual backbone angles~\cite{hamelryck2006sampling}. TorusDBN uses 55 hidden states with bivariate von Mises distributions for the backbone dihedral angles $\phi$ and $\psi$ and a categorical peptide-bond cis/trans variable~\cite{boomsma2008generative}. Our implementation uses the published BackboneDBN 0.24 model definitions and parameter files, retaining state counts, amino-acid emissions, angular families and chain dependencies while stabilizing hidden-path backtracking, integrating concentrated angular densities over their full support and checking atomic reconstruction across residue types. Complete-chain sampling is conditioned on the amino-acid sequence with secondary-structure variables marginalized. A secondary sensitivity analysis adapts both model families on the fitting conformations using weighted forward--backward expectation steps and numerical maximization of angular emission objectives; the FB5-HMM and TorusDBN adaptations stopped after 203 and 214 accepted updates, respectively, at the prescribed iteration budget.

The generator comparison in Table~\ref{tab:external-models} converts two independent batches of 512 complete conformations per test sequence into four-$\mathrm{C}_\alpha$ geometry and divides torsion into 36 equal intervals. Splitting each interval of the fitting-defined $5\times5$ bending grid at its midpoint yields 100 finer cells in which generated angular probabilities are estimated with a symmetric Dirichlet pseudocount of one half per torsion interval and mixed using the observed fine-cell proportions within each protein. Elastic densities are integrated over the same angular intervals at each observed bend pair, yielding a discrete probability score on shared bending support. Internal positions are weighted equally within each protein, followed by equal chains within a sequence group and equal groups, with 95\% confidence intervals from 2000 paired sequence-group resamples. Circular concentration within region $I_\sigma$ is $R_\sigma=|\langle e^{i\tau}\rangle_{\tau\in I_\sigma}|$, computed by dividing the joint complex moment by the probability of that region before taking its magnitude. For a generative model, the proper-rotation RMSD helicity criterion is applied to generated coordinates before bending standardization. Elastic predictions integrate each observed bond-length-specific mask $H_i(t)$ against its angular density.

Figure~\ref{fig:controlled}, Figure~\ref{fig:controlled-angles} and the complete-target elastic comparison use direct conditional evaluation of the published FB5-HMM on the complete observed targets (4000 saved units per ATLAS chain and every tenth frame at 0.1 ns intervals over $5<t\leq500$ ns in the controlled trajectories), where full-sequence amino-acid messages and the two observed bends determine state weights and numerical integration of the Kent components gives the continuous torsion density without discarding bending regions. Figure~\ref{fig:controlled-angles} displays the corresponding observed and predicted probabilities in 36 equal angular intervals. Within each trajectory, internal four-$\mathrm{C}_\alpha$ windows have equal weight, with equal selected valid frames within each window. The resulting trajectory distributions are averaged equally over the twelve trajectories of each system and protocol. A position-resolved sampling comparison reuses the two 512-conformation generator batches across all 1744 ATLAS test chains (retaining 94.00\% mean group-weighted support) and 72 controlled trajectories (57.99\% mean support), as summarized in Tables~\ref{tab:position-sampling} and \ref{tab:position-mc}.

The representative-window comparison in Figure~\ref{fig:external-comparison}b to d uses two independent batches of 65\,536 complete conformations per sequence and reference model. Each trajectory uses the same retained observations and observed bending-cell weights for all angular predictions. The single-state and two-state elastic densities are evaluated at the observed bend pairs, integrated with 1024 angular nodes and projected into the same 36 torsion intervals. The resulting distributions are averaged equally over the 24 trajectories of each system. The mean shared bending coverage is 99.63\%, 99.70\% and 100\% for Fs, $(\mathrm{AAQAA})_3$ and HP35, respectively. Each generated batch is averaged separately over the same trajectories before its pointwise range is displayed. Table~\ref{tab:local-mc} reports the corresponding sampling variation.

Standard Lifson--Roig theory is evaluated in residue-level variables~\cite{lifson1961theory,vitalis201250}, assigning a helical state when $\phi\in[-90^\circ,-30^\circ]$ and $\psi\in[-77^\circ,-17^\circ]$~\cite{sorin2005exploring} and a unit-weight coil state otherwise. A helical residue has weight $w_i$ when both neighbors are helical and $v_i$ otherwise, with exterior coil states, so a maximal helical run of length $n\geq2$ has homogeneous weight $v^2w^{n-2}$ and an isolated helical residue has weight $v$. Exact finite-chain transfer calculations give the partition function, occupancy, run-length and boundary probabilities, with 40 amino-acid-dependent log weights fitted by maximum likelihood on the 194 fitting chains and evaluated on 48 complete conformations per ATLAS test chain (Table~\ref{tab:standard-lr}) and across the controlled trajectories (Table~\ref{tab:controlled-lr}).

\subsection{Conditional Kent mixtures, moment projections and ablations}
\label{app:kent}
\label{app:reference-projection}

For a Kent state $s$ with orthonormal axes $\boldsymbol\gamma_{1s},\boldsymbol\gamma_{2s},\boldsymbol\gamma_{3s}$, concentration $\kappa_s$ and ovalness $\beta_s$, write $\mathbf h_s=\kappa_s\boldsymbol\gamma_{1s}$ and $Q_s=\beta_s(\boldsymbol\gamma_{2s}\boldsymbol\gamma_{2s}^{\mathsf T}-\boldsymbol\gamma_{3s}\boldsymbol\gamma_{3s}^{\mathsf T})$. With internal virtual bond angle $\theta=\pi-\kappa$, $c=\cos\theta$ and $u=\sin\theta$, substituting $\mathbf z=(c,u\cos\tau,u\sin\tau)$ into the spherical exponent $\mathbf h_s^{\mathsf T}\mathbf z+\mathbf z^{\mathsf T}Q_s\mathbf z$ gives the unnormalized angular log density
\begin{equation}
\label{eq:kent-conditional}
 \ell_s(\tau;\theta)=a_s\cos\tau+b_s\sin\tau+c_s\cos2\tau+d_s\sin2\tau,
\end{equation}
where the first-harmonic coefficients are $a_s=u h_{s2}+2cuQ_{s12}$ and $b_s=u h_{s3}+2cuQ_{s13}$, and the second-harmonic coefficients are
\begin{equation}
\label{eq:kent-cd}
 c_s=\tfrac12u^2(Q_{s22}-Q_{s33}),\qquad d_s=u^2Q_{s23}.
\end{equation}
The omitted constant and the sphere-area Jacobian $\sin\theta$ cancel in $p_s(\tau\mid\theta)=e^{\ell_s}/\int e^{\ell_s}dt$ while remaining in the bend marginal $m_s(\theta)$. Letting $q_{uv}(\mathbf a)$ denote the posterior probability of two adjacent hidden states given the amino-acid sequence $\mathbf a$, conditional independence of emissions expresses the conditional torsion density as the state mixture $p(\tau\mid\mathbf a,\theta_L,\theta_R)=\sum_v w_v p_v(\tau\mid\theta_R)$, with state weights
\begin{equation}
\label{eq:kent-mixture}
 w_v=\frac{m_v(\theta_R)\sum_u q_{uv}(\mathbf a)m_u(\theta_L)}
 {\sum_{u,k}q_{uk}(\mathbf a)m_u(\theta_L)m_k(\theta_R)}.
\end{equation}
Both bends influence the mixture weights, and the right bend also changes the component shapes. Equal components proportional to $e^{\pm a\cos\tau}$ give a mixture proportional to $\cosh(a\cos\tau)$, whose log density contains $-a^4\cos4\tau/96+O(a^6)$ for small $a$. A mixture therefore generally exceeds the two-harmonic log-density family even though each component belongs to it.

For a frozen FB5-HMM conditional density $f$ at one input, let $\mathbf T(t)=(\cos t,\sin t,\cos2t,\sin2t)$. Its information projection onto the local generalized von Mises family has coefficients $\bm\eta_*$ satisfying $\nabla\log Z(\bm\eta_*)=\mathbb E_f\mathbf T$, and moment matching gives
\begin{equation}
\label{eq:projection-identity}
 D(f\|p_{\bm\eta})=D(f\|p_{\bm\eta_*})+D(p_{\bm\eta_*}\|p_{\bm\eta}).
\end{equation}
The equality follows from integrating the affine difference of the two log densities against $f$ and $p_{\bm\eta_*}$, which share the same four moments. A global six-coefficient projection and two nested polynomial families of 16 coefficients on $(1,r,r^2,r^3)$ and 32 coefficients adding $(x_L,x_R,x_L^2,x_R^2)$ are fitted by convex optimization without penalty on the 9894 training inputs. The four frozen-model ablations retain or marginalize amino-acid emission factors and retain or remove adjacent-state dependence by replacing $q_{uv}(\mathbf a)$ with the product of its node marginals before conditioning on either bend.

\subsection{Gauge invariance and material-angle reduction}
\label{app:frames}
\label{app:elimination}

In the shared normal plane, $\mathbf b_L\cdot\mathbf b_R=\cos\tau$ and $\mathbf t\cdot(\mathbf b_L\times\mathbf b_R)=\sin\tau$, yielding Eq.~\eqref{eq:elastic-metrics} for the chosen generator metric. Whereas the generators in Eq.~\eqref{eq:elastic-metrics} describe minimal bending rotations between adjacent tangents, Eq.~\eqref{eq:frame-transport} gives the full frame transport including torsion. Under a change of local normal basis by $\alpha_i$, a normal coordinate $z_i$ and its transport link $W_i$ transform as $z_i'=e^{-i\alpha_i}z_i$ and $W_i'=e^{-i\alpha_{i+1}}W_ie^{i\alpha_i}$, preserving the covariant difference
\begin{equation}
\label{eq:si_material_link}
 |z_{i+1}'-W_i'z_i'|^2=|z_{i+1}-W_iz_i|^2.
\end{equation}
Setting the links to unity along the open chain gives the parallel-frame field and adjacent-difference identity in Eq.~\eqref{eq:hasimoto-mismatch}. Setting $W_i=e^{i\tau_0}$ with $\tau_0=\operatorname{atan2}(J_s,K)$ gives the linked difference $|z_{i+1}-e^{i\tau_0}z_i|^2=x_i^2+x_{i+1}^2-2x_ix_{i+1}\cos(\tau_i-\tau_0)$. With $\tilde K=\sqrt{K^2+J_s^2}$, the complete pair energy is
\begin{equation}
\label{eq:linked-pair-energy}
 E_{\mathrm{pair}}=\frac{\tilde K}{2}|z_{i+1}-e^{i\tau_0}z_i|^2
 +\frac{K-\tilde K}{2}(x_i^2+x_{i+1}^2).
\end{equation}
The second term cancels from a normalized torsion density at fixed bends. State-dependent bend constants must remain in the mixture weights, as in Eq.~\eqref{eq:state-elastic-density}. The SU(2) lifts $U_i$ and $-U_i$ describe the same vector rotation.

Let $\mathbf a_N$ point from C$_\alpha$ to N in the second residue of a window. For a nonzero projection with $P_\perp=I-\mathbf t\mathbf t^{\mathsf T}$, the unit normal projection $\mathbf m_N=P_\perp\mathbf a_N/|P_\perp\mathbf a_N|$ and left normal $\mathbf n_L=\mathbf b_L\times\mathbf t$ define the material angle
\begin{equation}
\label{eq:si_material_angle}
 d=\operatorname{atan2}(\mathbf m_N\cdot\mathbf b_L,\mathbf m_N\cdot\mathbf n_L).
\end{equation}
The right normal is $\mathbf n_R=\cos\tau\,\mathbf n_L+\sin\tau\,\mathbf b_L$, giving $\mathbf m_N\cdot\mathbf n_R=\cos(\tau-d)$. Rotating the fourth C$_\alpha$ around the shared tangent at fixed right bend changes $\tau$ while leaving $\mathbf m_N$ and $d$ unchanged. For Eq.~\eqref{eq:material-energy} at fixed bends $q$ and circular measure $d\tau\,dd$, defining $Z_m(q)=\int_{-\pi}^{\pi}e^{-\beta E_m(q,d)}\,dd$ and the uncoupled material weight $w_0(d\mid q)=Z_m(q)^{-1}e^{-\beta E_m(q,d)}$ yields the coupling factor
\begin{equation}
\label{eq:si_material_elimination}
 \Lambda(q,\tau)=\int_{-\pi}^{\pi}w_0(d\mid q)e^{\beta Jr\cos(\tau-d)}\,dd.
\end{equation}
Integrating out $d$ then gives the reduced conditional torsion density
\begin{equation}
\label{eq:si_material_posterior}
 p(\tau\mid q)=\frac{p_0(\tau\mid q)\Lambda(q,\tau)}
 {\int_{-\pi}^{\pi}p_0(t\mid q)\Lambda(q,t)\,dt}.
\end{equation}
The coupled marginal includes an additional factor $\mathcal M(q,d)=\int p_0(t\mid q)e^{\beta Jr\cos(t-d)}\,dt$, so $p(d\mid q)\propto w_0(d\mid q)\mathcal M(q,d)$. Writing $C_\tau=\cos(\tau-d)$, its uncoupled mean is $\mathbb E C_\tau=\operatorname{Re}(m_1^*e^{i\tau})$ and its uncoupled variance is
\begin{equation}
\label{eq:si_material_cumulant}
 \operatorname{Var}C_\tau=\frac12\{1-|m_1|^2+
 \operatorname{Re}[(m_2^*-(m_1^*)^2)e^{2i\tau}]\},
\end{equation}
whose first two cumulants give Eq.~\eqref{eq:material-cumulants}. For $Z=e^{id}$, the circular moments satisfy the bound
\begin{equation}
\label{eq:app_moment_constraint}
 |m_2-m_1^2|=|\mathbb E(Z-m_1)^2|
 \leq\mathbb E|Z-m_1|^2=1-|m_1|^2.
\end{equation}
For a von Mises weight $w_0(d)\propto e^{\eta\cos(d-\mu)}$, the integral evaluates to
\begin{equation}
 \Lambda(\tau)=\frac{I_0\!\left(\sqrt{\eta^2+z^2+2\eta z\cos(\tau-\mu)}\right)}{I_0(\eta)},
\label{eq:si_material_bessel}
\end{equation}
where $I_0$ is the modified Bessel function. For the effective angular free energy $F=-\beta^{-1}\log\int e^{-\beta E}\,dd$ and conditional material distribution $P_\tau\propto e^{-\beta E}$, differentiation at fixed $q$ gives $F'=\langle E'\rangle_{P_\tau}$ and
\begin{equation}
\label{eq:material-force-body}
 F''=\langle E''\rangle_{P_\tau}-\beta\operatorname{Var}_{P_\tau}(E'),
\end{equation}
where the fluctuation term reduces curvature relative to the conditional mean direct curvature.

\subsection{Atomic matching and material-subspace projection}
\label{app:controlled-matching}
\label{app:controlled-atomic-results}

For candidate four-residue window $j$, each of eight start-condition sources pools three repeats and supplies a two-bend histogram $N_{j,a}^{uv}$ with bin width 0.005. The selection score is
\begin{equation}
\label{eq:window-selection}
 S_j=\sum_{uv\,(x_Lx_R<0.23214)}\min_{a=1,\ldots,8}
 \frac{N_{j,a}^{uv}}{N_{j,a}},
\end{equation}
where $N_{j,a}$ counts valid frames for window $j$ in source $a$, and the largest score selects the window with the lowest residue index breaking ties. Matching pairs each negative query with its nearest nonnegative control in the maximum norm of $(x_L,x_R)$ at tolerances 0.005 and 0.002 with control reuse permitted. Backbone angles are obtained from saved coordinates~\cite{mcgibbon2015mdtraj}, a backbone hydrogen bond is occupied when the H--O distance is below 0.25 nm and the N--H--O angle exceeds $120^\circ$, and internal contacts use a 0.45 nm heavy-atom distance threshold between residues separated by at least six positions. Projecting the added material coefficient functions $(r\cos d,r\sin d,0,0)$ onto the six-function span separates two $L^2$ projections onto $\{1,r\}$. With the two-by-two Gram matrix $G_2$ having entries $1$, $\langle r\rangle$ and $\langle r^2\rangle$ and moment vectors $\mathbf u_c=(\langle r\cos d\rangle,\langle r^2\cos d\rangle)^{\mathsf T}$ and $\mathbf u_s=(\langle r\sin d\rangle,\langle r^2\sin d\rangle)^{\mathsf T}$, the relative squared-norm residual is
\begin{equation}
\label{eq:controlled-material-projection}
 \epsilon_d^2=1-\frac{1}{\langle r^2\rangle}\Bigl(\mathbf u_c^{\mathsf T}G_2^\dagger\mathbf u_c
 +\mathbf u_s^{\mathsf T}G_2^\dagger\mathbf u_s\Bigr),
\end{equation}
where moments are normalized within each trajectory and averaged equally across the twelve trajectories in a group (Tables~\ref{tab:controlled-matching} and \ref{tab:selected-turns}). The projection uses the measured coupled trajectory distribution, and $100(1-\epsilon_d^2)$ measures the fraction of coefficient-function squared $L^2$ norm within the existing span. It is an empirical projection and does not estimate the uncoupled moments of $w_0$. The dense matching analysis retains all 0.01 ns frames. Table~\ref{tab:selected-turns} instead uses the 0.1 ns common target and assigns negative probability by summing the 18 negative half-circle intervals. The dense HP35 single-state elastic probabilities are 32.96\% under AMBER and 31.20\% under CHARMM, while the common-target values are 33.00\% and 31.16\%.

\section*{Supporting Information}
\suppressfloats[t]

The Supporting Information retains the complete curvature controls, elastic variants, four-model common evaluation, reference ablations and moment projections, together with the starting structures, fixed coefficients, residue-state benchmarks, generator sensitivity and bend-matched atomic diagnostics. Figures~\ref{fig:external-comparison}, \ref{fig:controlled-angles} and \ref{fig:hp35-amber} present helical-run and representative-window predictions, full-system angular distributions and the AMBER HP35 turn, respectively.

\Needspace{13\baselineskip}
\section*{Data and Software Availability}

ATLAS trajectories and the earlier Fs archive are publicly available~\cite{vander2024atlas,fsarchive}. The 72 molecular dynamics trajectories generated in this study are available on Zenodo at \url{https://doi.org/10.5281/zenodo.23895045}. The simulation and analysis code, initial structures, and fitted model parameters are available on GitHub at \url{https://github.com/wyqmath/protein-backbone-elasticity}.

\section*{Author Contributions}

Yiquan Wang contributed to conceptualization, methodology, software, investigation, formal analysis, data curation, visualization, writing of the original draft, and manuscript review and editing. Mingyu Bian contributed to methodology, software, investigation, formal analysis, and visualization. Jiayuan Ma contributed to methodology, software, investigation, and formal analysis.

\FloatBarrier
{\fontsize{8.5}{9.5}\selectfont\justifying
\setlength{\baselineskip}{9.5pt}
\setlength{\bibsep}{2pt}
\bibliographystyle{unsrtnat}
\bibliography{references}
}
\clearpage
\onecolumn
\setcounter{figure}{0}
\setcounter{table}{0}
\renewcommand{\thefigure}{S\arabic{figure}}
\renewcommand{\thetable}{S\arabic{table}}
\fontsize{10.5}{12.8}\selectfont
\raggedbottom
\setlength{\floatsep}{14pt plus 2pt minus 2pt}
\setlength{\textfloatsep}{14pt plus 2pt minus 2pt}
\captionsetup{font=small,labelfont=bf,labelsep=period,justification=justified,singlelinecheck=false}
\captionsetup[figure]{name=Figure}
\makeatletter
\setlength{\@fptop}{0pt}
\setlength{\@fpsep}{14pt}
\setlength{\@fpbot}{0pt plus 1fil}
\makeatother

\section*{Supporting Information}

\subsection{Curvature controls and elastic variants}
\label{si:elastic-controls}

The curvature controls in Table~\ref{tab:atlas-models} use the same fitting observations and matched model sizes where applicable. The original continuous baseline loss decreases from 1.490 to 1.160 nats, with a paired reduction of 0.330 nats and 95\% interval $[0.327,0.334]$. The curvature-product model has a smaller advantage over the sum control, with paired loss difference $-0.011$ nats and interval $[-0.012,-0.011]$. Equation~\eqref{eq:product-sum} explains their shared first-order bend response.

\begin{table}[htbp]
\centering\small
\caption{Performance on 1744 ATLAS test chains. Entries are group-weighted means with 95\% group-bootstrap intervals. Direction and helix errors are absolute chain-fraction errors for negative torsion and geometric helicity, respectively, in percentage points.}
\label{tab:atlas-models}
\begin{tabularx}{\textwidth}{@{}lr*{3}{>{\centering\arraybackslash}X}@{}}
\toprule
Model & Coefficients & Negative log density & Direction error & Helix error\\
\midrule
Curvature-independent control & 4 & \shortstack{1.490\\{\scriptsize(1.476, 1.504)}} & \shortstack{12.26\\{\scriptsize(11.80, 12.73)}} & \shortstack{21.84\\{\scriptsize(21.20, 22.50)}} \\
Single-state elastic model & 6 & \shortstack{1.160\\{\scriptsize(1.143, 1.176)}} & \shortstack{2.15\\{\scriptsize(2.06, 2.24)}} & \shortstack{10.41\\{\scriptsize(10.06, 10.76)}} \\
Bend-sum control & 6 & \shortstack{1.171\\{\scriptsize(1.155, 1.187)}} & \shortstack{2.24\\{\scriptsize(2.15, 2.34)}} & \shortstack{10.99\\{\scriptsize(10.62, 11.35)}} \\
Minimum-bend control & 6 & \shortstack{1.209\\{\scriptsize(1.192, 1.226)}} & \shortstack{2.77\\{\scriptsize(2.65, 2.89)}} & \shortstack{11.15\\{\scriptsize(10.78, 11.52)}} \\
\bottomrule
\end{tabularx}
\end{table}

\begin{table}[htbp]
\centering\small
\caption{Elastic extensions and published FB5-HMM on the complete observed targets. Sizes denote elastic coefficients or reference hidden states. Density losses are continuous per-radian scores. Helix errors are absolute chain or trajectory fraction errors in percentage points. ATLAS means give equal weight to sequence groups; controlled means give equal weight to the 24 system, protocol and start combinations, each with three repeats.}
\label{tab:elastic-upgrades}
\begin{tabularx}{\textwidth}{@{}l r *{4}{>{\centering\arraybackslash}X}@{}}
\toprule
 & & \multicolumn{2}{c}{ATLAS} & \multicolumn{2}{c}{Controlled trajectories}\\
Model & Size & Density loss & Helix error & Density loss & Helix error\\
\midrule
Single-state elastic model & 6 & 1.160 & 10.41 & 1.231 & 10.73\\
Anisotropic elastic model & 8 & 1.056 & 6.16 & 1.156 & 7.11\\
Unconstrained anisotropic control & 8 & 1.016 & 4.70 & 1.153 & 5.87\\
Linear second-harmonic control & 8 & 1.018 & 3.74 & 1.144 & 5.11\\
Two-state elastic model & 20 & 0.790 & 1.65 & 0.892 & 2.87\\
Sequence-associated two-state model & 40 & 0.780 & 1.40 & 0.893 & 2.86\\
FB5-HMM, published & 75 states & 0.448 & 1.56 & 0.658 & 3.34\\
\bottomrule
\end{tabularx}
\end{table}

The constrained anisotropic extension lowers continuous ATLAS loss from 1.160 to 1.056 nats and absolute chain helix-fraction error from 10.41 to 6.16 percentage points. The equal-size unconstrained and linear second-harmonic controls give lower losses, preserving a predictive cost of the positive-stiffness restriction. The two-state elastic model gives 0.790 nats and 1.65 points. Its sequence-associated variant gives 0.780 nats and 1.40 points, with helix Brier score 0.0404 against 0.0432 for the local model. Sign-fraction errors remain near two points. All variants are fitted to torsions, with geometric helix labels used only for evaluation. On the controlled trajectories, the sequence-associated variant gives loss 0.893 nats, helix-fraction error 2.86 points and helix Brier score 0.0508, compared with 0.892, 2.87 and 0.0545 for the local model. Direct FB5-HMM gives helix Brier score 0.0484. On the controlled trajectories, the sequence-associated variant slightly increases angular loss while reducing helix-fraction error and the helix Brier score.

\subsection{Common angular comparison and alternative error aggregation}
\label{si:common-evaluation}

Table~\ref{tab:common-four-models} records the four-model comparison used in Figure~\ref{fig:upgrade-validation}. The chain-fraction error sums supported probabilities within each chain before taking the absolute difference. Table~\ref{tab:external-models} retains the earlier cellwise absolute direction and helix errors, total variation and bootstrap intervals. Its absolute errors are averaged over bending cells before cohort aggregation and cannot be read as chain-fraction errors.

\begin{table}[htbp]
\centering\small
\caption{Four-model common-support evaluation on 1744 ATLAS chains in 1317 sequence groups. Angular loss uses 36 torsion intervals, and helix error is the absolute chain-fraction difference in percentage points. Parentheses give 95\% limits from 2000 paired group bootstrap resamples. All models use the same 100 bending cells and observed weights.}
\label{tab:common-four-models}
\begin{tabularx}{\textwidth}{@{}Xll@{}}
\toprule
Model & Angular loss (nat) & Chain helix error\\
\midrule
Single-state elastic model & 2.899 (2.884, 2.915) & 10.43 (10.07, 10.78)\\
Two-state elastic model & 2.581 (2.558, 2.604) & 1.65 (1.58, 1.72)\\
FB5-HMM, published & 2.444 (2.423, 2.465) & 1.64 (1.57, 1.72)\\
TorusDBN, published & 2.565 (2.547, 2.582) & 3.81 (3.68, 3.94)\\
\bottomrule
\end{tabularx}
\end{table}

\begin{table}[tp]
\centering\small
\caption{Common conditional comparison on all 1744 ATLAS test chains. Angular loss is the negative log probability of a torsion interval. TV is total variation between the 36-interval distributions. Direction and helix entries are cellwise absolute errors in percentage points. Entries give group-weighted means with 95\% group-bootstrap intervals on shared bending support.}
\label{tab:external-models}
\begin{tabularx}{\textwidth}{@{}l*{4}{>{\centering\arraybackslash}X}@{}}
\toprule
Model & Angular loss (nat) & TV & Direction error & Helix error\\
\midrule
Single-state elastic model & \shortstack{2.899\\{\scriptsize(2.885, 2.914)}} & \shortstack{0.471\\{\scriptsize(0.468, 0.474)}} & \shortstack{14.50\\{\scriptsize(14.32, 14.70)}} & \shortstack{11.23\\{\scriptsize(10.90, 11.55)}} \\
FB5-HMM, published & \shortstack{2.444\\{\scriptsize(2.424, 2.464)}} & \shortstack{0.222\\{\scriptsize(0.220, 0.224)}} & \shortstack{5.05\\{\scriptsize(4.96, 5.14)}} & \shortstack{2.69\\{\scriptsize(2.63, 2.75)}} \\
TorusDBN, published & \shortstack{2.565\\{\scriptsize(2.548, 2.582)}} & \shortstack{0.294\\{\scriptsize(0.292, 0.296)}} & \shortstack{6.15\\{\scriptsize(6.06, 6.24)}} & \shortstack{4.94\\{\scriptsize(4.83, 5.05)}} \\
\bottomrule
\end{tabularx}
\end{table}

Applying Eq.~\eqref{eq:loss-decomposition} to the 36-interval probabilities gives direction loss 0.436 nats and within-direction interval loss 2.463 nats for the single-state elastic model. Published FB5-HMM reduces these components by 0.088 and 0.367 nats, and published TorusDBN by 0.078 and 0.256 nats. Most of the reference gain in this decomposition comes from within-direction shape, with an additional improvement in directional allocation. Finite-budget adaptation improves FB5-HMM interval loss to 2.416 nats and its cellwise geometric helix error to 1.91 percentage points. TorusDBN adaptation reduces interval loss to 2.548 nats and improves direction allocation while increasing cellwise geometric helix error to 5.09 points, indicating that angular-loss and structural-error changes need not coincide. The earlier controlled generator comparison has interval losses 2.600 and 2.679 nats for published FB5-HMM and TorusDBN, against 2.948 for the single-state elastic model, with mean shared bending coverage 93.85\% for Fs, 91.21\% for $(\mathrm{AAQAA})_3$ and 99.87\% for HP35 (Table~\ref{tab:controlled-external}).

\subsection{Reference ablations and circular-moment projections}
\label{si:reference-diagnostics}

The continuous reference diagnostics use only the 9894 fitting inputs and leave the published FB5-HMM parameters fixed. Its full conditional loss is 0.430 nats. Marginalizing amino-acid information adds 0.167 nats, removing adjacent-state dependence adds 0.283 nats, and removing both gives 0.922 nats. The nonadditive interaction is 0.043 nats, and Eqs.~\eqref{eq:kent-conditional} and \eqref{eq:kent-mixture} describe how sequence and adjacent-state weights enter the reference prediction.

Matching the first two circular harmonics of the reference independently at each input leaves mean divergence 0.139 nats. Restricting the coefficients to the global six-coefficient curvature-product form adds 0.609 nats, even when optimized against the reference moments. Replacing the original coefficients by this optimum changes divergence by only 0.007 nats. Within nested coefficient spaces, a nonlinear product response removes 0.203 nats and retaining the separate bends removes another 0.117 nats. The restriction therefore involves bend response and contextual state mixing as well as angular structure outside the two-harmonic family. Equation~\eqref{eq:projection-identity} gives the exact information-projection decomposition used in this comparison.

\subsection{Simulation inputs and supplementary benchmarks}

\begin{table}[htbp]
\centering\small
\caption{Four starting conformations per system. Start numbers follow the order shown. Each start has three repeats under both simulation conditions.}
\label{tab:data}
\begin{tabularx}{\textwidth}{@{}lrX@{}}
\toprule
System & Residues & Starting structures\\
\midrule
Fs & 21 & Archive-derived conformations with initial hydrogen-bond-based DSSP~\cite{kabsch1983dictionary} helicities of 0\%, 61.9\%, 90.5\% and 33.3\%\\
$(\mathrm{AAQAA})_3$ & 15 & Regular helix, local relaxation, central bend and extended chain\\
HP35 N27H & 35 & Crystal structure (PDB ID 1YRF) and deterministic backbone perturbations of $1^\circ$, $1.5^\circ$ and $2^\circ$\\
\bottomrule
\end{tabularx}
\end{table}

\begin{table}[htbp]
\centering\small
\caption{Coefficients of the fixed ATLAS densities in Eq.~\eqref{eq:angular-density}. Columns give $a_0,b_0,c_0,d_0,a_1,b_1$; values are rounded to four decimal places, and full precision is retained in the parameter files.}
\label{tab:fixed-coefficients}
\begin{tabular}{@{}lrrrrrr@{}}
\toprule
Model & $a_0$ & $b_0$ & $c_0$ & $d_0$ & $a_1$ & $b_1$\\
\midrule
Curvature-independent control & $-0.1205$ & $0.4468$ & $-0.1021$ & $1.3587$ & $0$ & $0$\\
Single-state elastic model & $-2.4096$ & $-2.2665$ & $-0.0494$ & $1.0411$ & $13.8967$ & $16.4633$\\
Bend-sum control & $-4.7553$ & $-4.8008$ & $-0.0626$ & $1.0539$ & $5.7119$ & $6.4687$\\
Minimum-bend control & $-3.1141$ & $-2.9137$ & $-0.0726$ & $1.1032$ & $8.1433$ & $9.1412$\\
\bottomrule
\end{tabular}
\end{table}

\begin{figure}[tp]
\centering
\includegraphics[width=.98\textwidth]{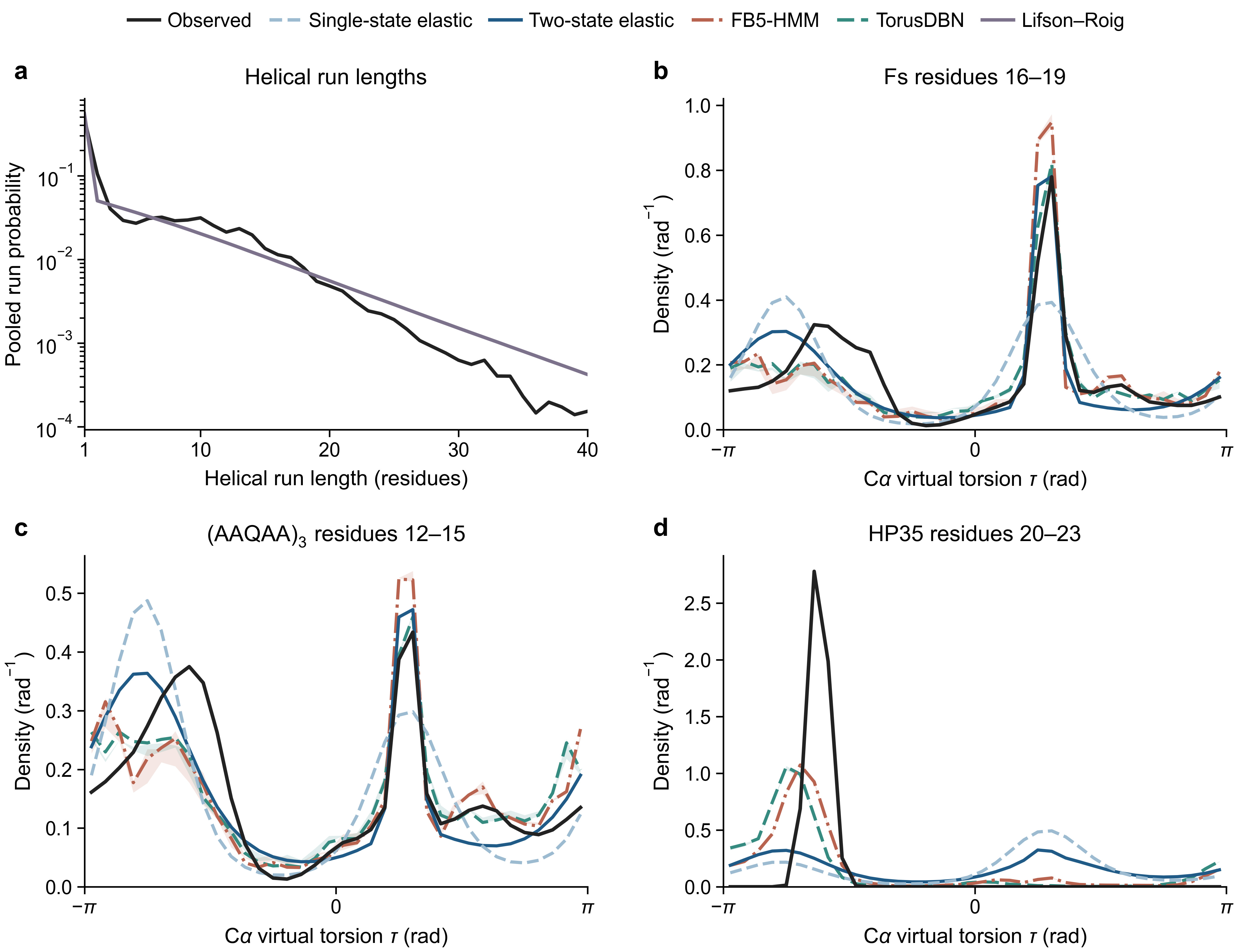}
\caption{Helical-run and representative-window predictions. Panel a compares observed and standard Lifson--Roig run-length probabilities on 1744 ATLAS test chains in 1317 sequence groups, using the atomic $\phi/\psi$ definition of residue helicity in Methods. Mean run counts are pooled with equal sequence-group weights before normalization. Lengths 1 to 40 are shown without renormalizing this range. Panels b to d show the C$_\alpha$ virtual-torsion distributions for Fs residues 16--19, $(\mathrm{AAQAA})_3$ residues 12--15 and HP35 residues 20--23, respectively. The observed, single-state elastic, two-state elastic, FB5-HMM and TorusDBN curves use the same shared bending support and observed bending-cell weights within each trajectory, followed by equal averaging over the 24 trajectories of each system. Mean retained observations are 99.63\%, 99.70\% and 100\%, respectively. Lines join the centers of 36 equal angular intervals, with density per radian. Shading for FB5-HMM and TorusDBN spans the predictions from two independent batches of 65\,536 complete sequence conformations after each batch is averaged over the same trajectories. Table~\ref{tab:local-mc} reports the sampling variation.}
\label{fig:external-comparison}
\end{figure}

\begin{figure}[tp]
\centering
\includegraphics[width=.96\textwidth,height=.72\textheight,keepaspectratio]{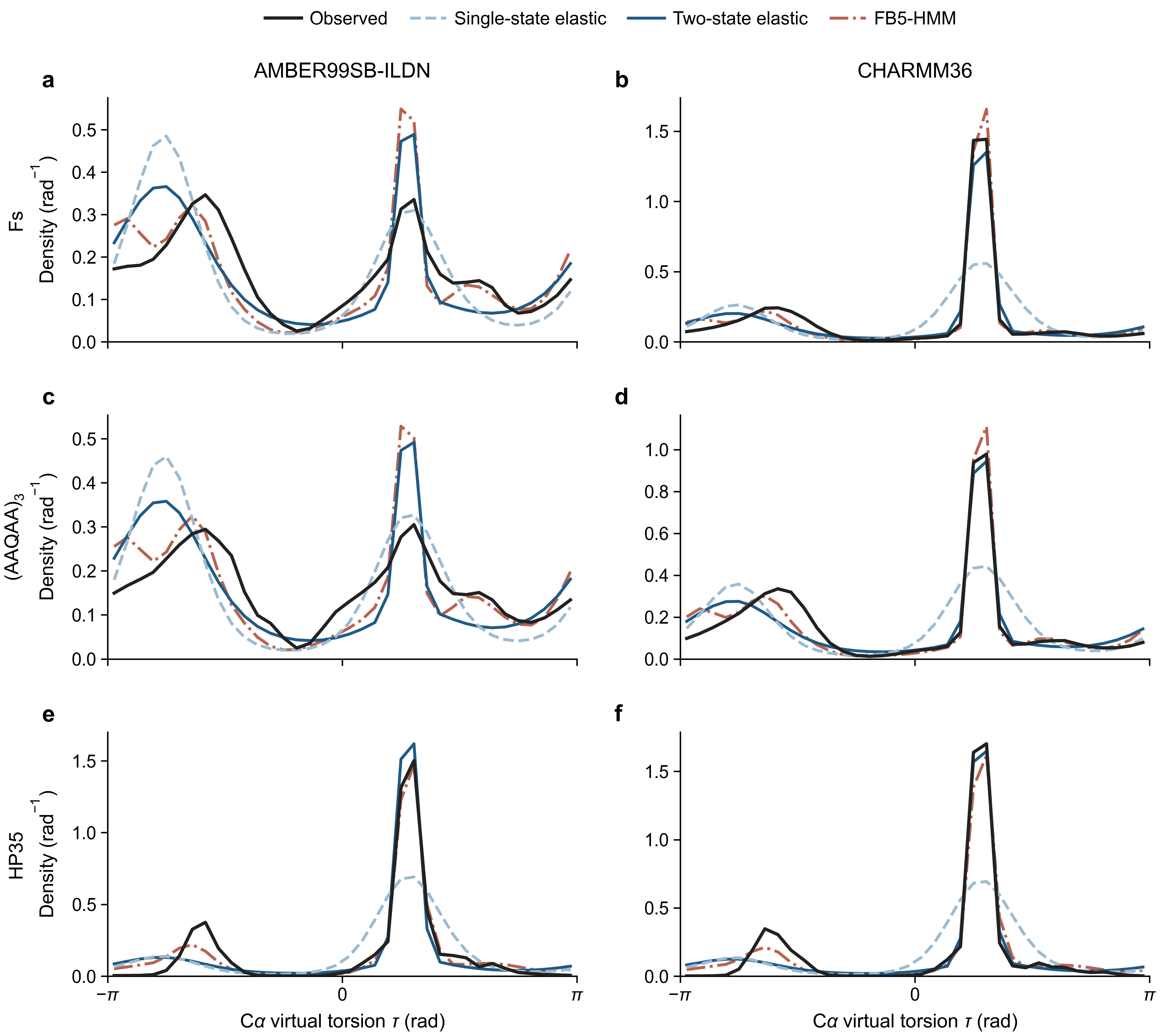}
\caption{Complete angular distributions across the controlled systems and force-field protocols. Rows show Fs, $(\mathrm{AAQAA})_3$ and HP35, and columns show AMBER99SB-ILDN and CHARMM36m. Each panel compares observations with the single-state elastic, two-state elastic and FB5-HMM predictions. The elastic parameters are fixed from ATLAS, and FB5-HMM uses its published parameters. The curves include all internal four-C$_\alpha$ windows and selected frames at 0.1 ns intervals over $5<t\leq500$ ns. Each window has equal weight within a trajectory, with equal valid selected frames within each window, followed by equal averaging over twelve trajectories per system and protocol. Lines join the centers of the same 36 angular intervals for all curves, with density per radian.}
\label{fig:controlled-angles}
\end{figure}

\FloatBarrier
\begin{table}[htbp]
\centering\small
\caption{Standard residue-level Lifson--Roig predictions on 1744 test chains. Errors compare the identical atomic $\phi/\psi$ state definition. Means and 95\% intervals use equal sequence-group weights. The binary-state log loss is evaluated in residue-state probability mass.}
\label{tab:standard-lr}
\begin{tabularx}{\textwidth}{@{}Xrrl@{}}
\toprule
Observable & Mean & 95\% interval & Unit\\
\midrule
Absolute chain helix-fraction error & 11.23 & (10.77, 11.67) & percentage points \\
Residue occupancy error & 34.01 & (33.74, 34.27) & percentage points \\
Helix-start probability error & 12.29 & (12.21, 12.38) & percentage points \\
Helix-end probability error & 11.44 & (11.36, 11.52) & percentage points \\
Mean chain run-length TV & 0.275 & (0.270, 0.281) & probability \\
Binary-state log loss & 0.382 & (0.379, 0.386) & nat per observed residue \\
\bottomrule
\end{tabularx}
\end{table}

\FloatBarrier
\begin{table}[H]
\centering\small
\caption{Common conditional comparison across all internal units of the 72 controlled trajectories. Entries are equal-design-cell means with standard errors from the three molecular dynamics repeats per cell. Direction and helix errors are in percentage points on shared bending support. The last column separately gives half the absolute angular-loss difference between the two generator batches, in nat; its two-batch estimate has one degree of freedom.}
\label{tab:controlled-external}
\begin{tabularx}{\textwidth}{@{}l*{3}{>{\centering\arraybackslash}X}r@{}}
\toprule
Model & Angular loss (nat) & Direction error & Helix error & Generator batch variation\\
\midrule
Curvature-independent control & $3.247\pm0.020$ & $37.41\pm0.27$ & $22.81\pm0.80$ & n/a \\
Single-state elastic model & $2.948\pm0.022$ & $16.25\pm0.25$ & $11.74\pm0.47$ & n/a \\
Bend-sum control & $2.946\pm0.021$ & $16.33\pm0.22$ & $12.30\pm0.48$ & n/a \\
Minimum-bend control & $2.992\pm0.023$ & $18.62\pm0.30$ & $12.43\pm0.49$ & n/a \\
FB5-HMM, published & $2.600\pm0.035$ & $9.67\pm0.32$ & $5.28\pm0.29$ & 0.0003 \\
FB5-HMM, adapted & $2.505\pm0.032$ & $7.95\pm0.33$ & $5.25\pm0.29$ & 0.0016 \\
TorusDBN, published & $2.679\pm0.029$ & $9.67\pm0.35$ & $6.44\pm0.28$ & 0.0045 \\
TorusDBN, adapted & $2.606\pm0.029$ & $8.02\pm0.32$ & $5.28\pm0.29$ & 0.0064 \\
\bottomrule
\end{tabularx}
\end{table}

\begin{table}[H]
\centering\small
\caption{Standard Lifson--Roig residue-state predictions in the controlled systems over $5<t\leq500$ ns. Observed and predicted helical fractions are percentages. Residue occupancy errors are in percentage points. Each entry weights four starts equally and three repeats within each start. The same fixed sequence parameters are used in both protocols.}
\label{tab:controlled-lr}
\begin{tabular}{@{}llrrr@{}}
\toprule
System & Protocol & Observed helical fraction & Predicted helical fraction & Residue error\\
\midrule
$(\mathrm{AAQAA})_3$ & AMBER & 11.14 & 79.36 & 68.23 \\
$(\mathrm{AAQAA})_3$ & CHARMM & 36.56 & 79.36 & 44.70 \\
Fs & AMBER & 13.47 & 85.18 & 71.70 \\
Fs & CHARMM & 50.38 & 85.18 & 37.42 \\
HP35 & AMBER & 55.75 & 41.76 & 33.68 \\
HP35 & CHARMM & 66.18 & 41.76 & 40.91 \\
\bottomrule
\end{tabular}
\end{table}

\begin{table}[H]
\centering\small
\caption{Largest two-batch standard error among the 36 angular-bin probabilities at each representative window, in percentage points. Each of the two independent batches contains 65\,536 complete sequence conformations. Each batch curve is averaged over the same 24 molecular dynamics trajectories before its contrast is computed, leaving one degree of freedom for the sampling estimate.}
\label{tab:local-mc}
\begin{tabular}{@{}lrrrr@{}}
\toprule
System & Published FB5-HMM & Adapted FB5-HMM & Published TorusDBN & Adapted TorusDBN\\
\midrule
Fs & 0.42 & 0.79 & 0.48 & 0.62 \\
$(\mathrm{AAQAA})_3$ & 0.52 & 0.38 & 0.35 & 0.43 \\
HP35 & 0.28 & 0.48 & 0.32 & 0.36 \\
\bottomrule
\end{tabular}
\end{table}

\begin{table}[htbp]
\centering\small
\caption{Position-resolved sampling comparison over the complete test cohorts. All models use the same positions and observed bond lengths on common generated bending support. Losses score the 36 torsion intervals. Brier scores refer to the local helix label, and helix-fraction errors are in percentage points. External estimates reuse two batches of 512 complete conformations per sequence and model.}
\label{tab:position-sampling}
\begin{tabularx}{\textwidth}{@{}Xrrrrrr@{}}
\toprule
 & \multicolumn{3}{c}{ATLAS} & \multicolumn{3}{c}{Controlled trajectories}\\
Model & Loss & Brier & Helix error & Loss & Brier & Helix error\\
\midrule
Single-state elastic model & 2.877 & 0.0712 & 11.20 & 2.763 & 0.1221 & 16.28\\
Linear second-harmonic control & 2.731 & 0.0460 & 4.35 & 2.700 & 0.0997 & 9.37\\
Unconstrained anisotropic control & 2.730 & 0.0481 & 5.35 & 2.689 & 0.1010 & 10.14\\
Anisotropic elastic model & 2.772 & 0.0525 & 6.85 & 2.705 & 0.1037 & 11.49\\
Two-state elastic model & 2.530 & 0.0407 & 1.61 & 2.447 & 0.1033 & 7.47\\
Sequence-associated two-state model & 2.519 & 0.0379 & 1.36 & 2.479 & 0.1017 & 7.78\\
FB5-HMM, sampled & 2.545 & 0.0729 & 12.39 & 2.444 & 0.1091 & 13.40\\
TorusDBN, sampled & 2.694 & 0.0898 & 14.87 & 2.596 & 0.1314 & 17.73\\
\bottomrule
\end{tabularx}
\end{table}

\begin{table}[htbp]
\centering\small
\caption{Sampling sensitivity on the fixed common position-by-bend support. Losses are 36-interval scores from each separate batch of 512 conformations, with the same one-half pseudocount per bin applied separately to each batch. The final column averages the absolute half-difference of the two losses over members using the cohort weights. It measures sampling sensitivity and is separate from molecular dynamics repeat error. Pooled estimates in Table~\ref{tab:position-sampling} combine counts before regularization.}
\label{tab:position-mc}
\begin{tabular}{@{}llrrr@{}}
\toprule
Cohort & Model & Batch 1 loss & Batch 2 loss & Mean absolute half-difference\\
\midrule
ATLAS & FB5-HMM & 2.6981 & 2.6980 & 0.0071\\
ATLAS & TorusDBN & 2.8398 & 2.8387 & 0.0071\\
Fs & FB5-HMM & 2.5311 & 2.5395 & 0.0053\\
Fs & TorusDBN & 2.7683 & 2.7468 & 0.0120\\
$(\mathrm{AAQAA})_3$ & FB5-HMM & 2.7544 & 2.7505 & 0.0055\\
$(\mathrm{AAQAA})_3$ & TorusDBN & 2.9266 & 2.8862 & 0.0202\\
HP35 & FB5-HMM & 2.3138 & 2.2921 & 0.0109\\
HP35 & TorusDBN & 2.4993 & 2.5233 & 0.0120\\
Controlled, pooled & FB5-HMM & 2.5331 & 2.5274 & 0.0072\\
Controlled, pooled & TorusDBN & 2.7314 & 2.7188 & 0.0147\\
\bottomrule
\end{tabular}
\end{table}

\subsection{Atomic diagnostics and turn probabilities}

The dense CHARMM Fs ensemble has 53.66\% observed local helicity and 37.63\% under the single-state elastic model despite close direction allocation. At Fs residues 16--19, dense negative-torsion fractions are 37.18\% observed and 36.86\% predicted, while geometric helicities are 44.84\% and 32.90\%. Native-coordinate sampling for HP35 residues 20--23 assigns 86.26\% to 86.77\% negative torsion with published FB5-HMM and 87.52\% to 88.74\% with published TorusDBN, using all observed bending mass. These sampled estimates and angular peaks appear in Figure~\ref{fig:external-comparison}d. The reference curves in that figure estimate conditional distributions from sampled complete conformations, whereas Table~\ref{tab:selected-turns} evaluates FB5-HMM directly at every observed bend pair using full-sequence amino-acid messages.

\begin{table}[htbp]
\centering\small
\caption{Matching at tolerance 0.002 and material-direction diagnostics. Controls may be reused. HP35 has 594000 negative observations per condition and no nonnegative control. $R_d$ is material-angle concentration and $100\epsilon_d^2$ is the coefficient-function squared-norm residual.}
\label{tab:controlled-matching}
\label{tab:controlled-material}
\begin{tabular}{@{}llrrrrrr@{}}
\toprule
System & Condition & Window & Queries & Controls & Median $\Delta\tau$ & $R_d$ & $100\epsilon_d^2$\\
\midrule
Fs & AMBER & 16--19 & 288062 & 76260 & $105.0^\circ$ & 0.986 & 3.33\\
Fs & CHARMM & 16--19 & 198292 & 43813 & $118.2^\circ$ & 0.995 & 0.73\\
$(\mathrm{AAQAA})_3$ & AMBER & 12--15 & 321462 & 72873 & $92.7^\circ$ & 0.993 & 1.74\\
$(\mathrm{AAQAA})_3$ & CHARMM & 12--15 & 278030 & 42104 & $105.4^\circ$ & 0.992 & 1.28\\
HP35 & AMBER & 20--23 & 0 & 0 & n/a & 0.997 & 0.38\\
HP35 & CHARMM & 20--23 & 0 & 0 & n/a & 0.996 & 0.69\\
\bottomrule
\end{tabular}
\end{table}

\begin{table}[htbp]
\centering\small
\caption{Negative-torsion probabilities at the representative four-residue units, in percent. Predictions use all observed geometries from the matched 0.1 ns target over $5<t\leq500$ ns, averaged equally over twelve trajectories per system and protocol. Each external entry is direct conditional FB5-HMM evaluation.}
\label{tab:selected-turns}
\begin{tabularx}{\textwidth}{@{}llrr*{3}{>{\centering\arraybackslash}X}@{}}
\toprule
System & Protocol & Observed & Single-state elastic & Two-state elastic & Sequence-associated & FB5-HMM\\
\midrule
Fs & AMBER & 56.10 & 60.86 & 58.89 & 58.66 & 47.06\\
Fs & CHARMM & 37.21 & 36.87 & 36.00 & 35.38 & 31.68\\
$(\mathrm{AAQAA})_3$ & AMBER & 58.49 & 64.48 & 63.20 & 62.14 & 53.17\\
$(\mathrm{AAQAA})_3$ & CHARMM & 49.93 & 50.62 & 49.85 & 48.69 & 43.10\\
HP35 & AMBER & 100.00 & 33.00 & 49.35 & 49.56 & 87.67\\
HP35 & CHARMM & 100.00 & 31.16 & 48.02 & 48.23 & 87.26\\
\bottomrule
\end{tabularx}
\end{table}

\begin{figure}[htbp]
\centering
\includegraphics[width=.90\textwidth]{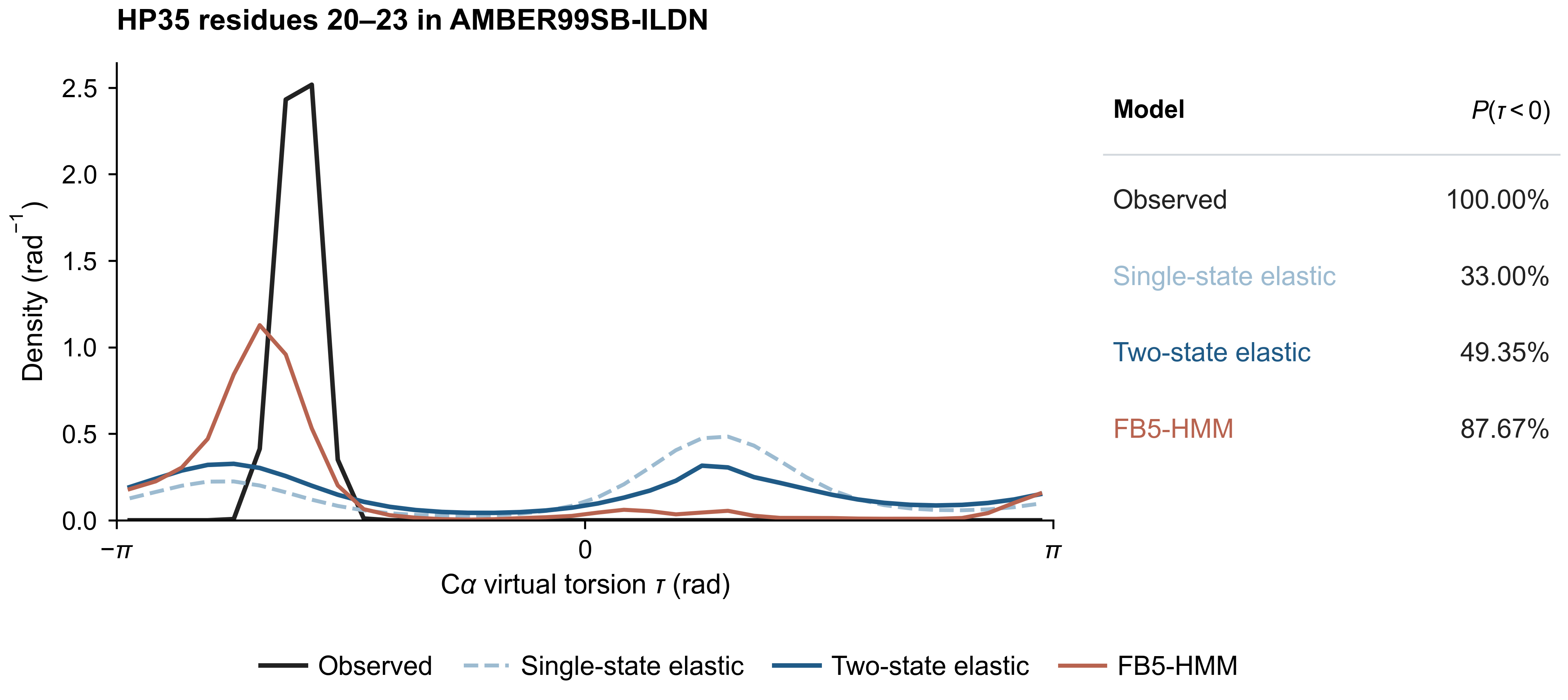}
\caption{Observed and predicted torsion distributions for HP35 residues 20--23 under AMBER99SB-ILDN. The elastic parameters are fixed from ATLAS, FB5-HMM uses its published parameters, and model colors match Figure~\ref{fig:matched-atomic}f. Probabilities are averaged equally over four starts and three repeats, using all observed geometries at 0.1 ns intervals over $5<t\leq500$ ns. The table gives the integrated negative-torsion probabilities. Curves join the centers of 36 angular intervals, with density per radian.}
\label{fig:hp35-amber}
\end{figure}

\end{document}